\documentclass[10pt,journal,compsoc]{IEEEtran}

\ifCLASSOPTIONcompsoc
  \usepackage[nocompress]{cite}
\else
  \usepackage{cite}
\fi

\ifCLASSINFOpdf
  \usepackage[pdftex]{graphicx}
\else
\fi

\usepackage{mdwtab}

\usepackage{newtxmath}
\usepackage{wasysym}
\usepackage{xcolor}

\makeatletter
\def\@cite#1#2{[{#1\if@tempswa ,~#2\fi}]}
\makeatother

\newcommand{\umlPageLink}[2]{\href{https://www.omg.org/spec/UML/2.5.1/PDF\#page=#1}{#2}}
\newcommand{\fumlPageLink}[2]{\href{https://www.omg.org/spec/FUML/1.5/PDF\#page=#1}{#2}}
\newcommand{\pssmPageLink}[2]{\href{https://www.omg.org/spec/PSSM/1.0/PDF\#page=#1}{#2}}
\newcommand{\umlPageCite}[2]{\cite[\protect\umlPageLink{#1}{#2}]{omg-uml}}
\newcommand{\fumlPageCite}[2]{\cite[\protect\fumlPageLink{#1}{#2}]{omg-fuml}}
\newcommand{\pssmPageCite}[2]{\cite[\protect\pssmPageLink{#1}{#2}]{omg-pssm}}

\newcommand{\umlClass}[1]{\texttt{#1}}
\newcommand{\stateSM}[1]{\textsf{#1}}
\newcommand{\transition}[1]{\textsf{#1}}
\newcommand{\event}[1]{\textsf{#1}}
\newcommand{\elementName}[1]{\textsf{#1}}

\newcommand{\inlineCmd}[1]{\texttt{#1}}

\newcommand{\guidelineSect}[1]{\subsubsection{#1}}
\newcommand{\guidelineSubsect}[1]{\textbf{#1:}}

\newcommand*{\warnNoText}{empty cell}
\newcommand*{\warnNo}{}
\newcommand*{\warnSlightly}{\Circle}
\newcommand*{\warnApplicable}{\RIGHTcircle}
\newcommand*{\warnImportant}{\CIRCLE}

\newcommand{\plantumlscale}{0.44}
\newcommand{\smMicroStepsPptScale}{.33}

\usepackage{enumitem}
\newlist{warnEnum}{enumerate}{1}
\newcommand{\warnLabelText}[1]{I#1)}
\newcommand{\warnRefText}[1]{I#1}
\newcounter{warnEnumiSaved}

\setlist[warnEnum]{label=\warnLabelText{\arabic*}, ref=\warnRefText{\arabic*}, resume}

\usepackage{hyperref}
\usepackage[nameinlink]{cleveref}

\crefname{figure}{\figurename}{Figs.}
\Crefname{figure}{\figurename}{Figs.}
\crefname{section}{Section}{Sections} %
\crefname{warnEnumi}{issue}{issues}

\usepackage{listings}
\usepackage[shortcuts]{extdash} %
\usepackage{makecell}
\usepackage{subcaption}
\usepackage{multirow}
\usepackage{url}

\begin{document}

\newcommand*{\titleText}{To Do or Not to Do: Semantics and Patterns for Do Activities in UML PSSM State Machines}
\newcommand{\copyrighttext}{\copyright{} 2024 IEEE. Personal use of this material is permitted. Permission from IEEE must be obtained for all other uses, in any current or future media, including reprinting/\\republishing this material for advertising or promotional purposes, creating new collective works, for resale or redistribution to servers or lists, or reuse of any copyrighted\\component of this work in other works.
In IEEE Transactions on Software Engineering. DOI: \href{https://doi.org/10.1109/TSE.2024.3422845}{10.1109/TSE.2024.3422845}}

\title{\titleText}

\author{M\'{a}rton~Elekes, Vince~Moln\'{a}r, and~Zolt\'{a}n~Micskei%
\IEEEcompsocitemizethanks{\IEEEcompsocthanksitem All authors are with the Department of Artificial Intelligence and Systems Engineering, Budapest University of Technology and Economics.\protect\\
E-mail: \{elekes,molnarv,micskei\}@mit.bme.hu
}%
}

\markboth{\titleText}{\titleText}
\IEEEpubid{\begin{tabular}[t]{@{}l@{}}\copyrighttext\end{tabular}}

\IEEEtitleabstractindextext{%
\begin{abstract}
State machines are used in engineering many types of software-intensive systems. UML State Machines extend simple finite state machines with powerful constructs. Among the many extensions, there is one seemingly simple and innocent language construct that fundamentally changes state machines' reactive model of computation: doActivity behaviors. DoActivity behaviors describe behavior that is executed independently from the state machine once entered in a given state, typically modeling complex computation or communication as background tasks. However, the UML specification or textbooks are vague about how the doActivity behavior construct should be appropriately used. This lack of guidance is a severe issue as, when improperly used, doActivities can cause concurrent, non-deterministic bugs that are especially challenging to find and could ruin a seemingly correct software design. The Precise Semantics of UML State Machines (PSSM) specification introduced detailed operational semantics for state machines. To the best of our knowledge, there is no rigorous review yet of doActivity's semantics as specified in PSSM. We analyzed the semantics by collecting evidence from cross-checking the text of the specification, its semantic model and executable test cases, and the simulators supporting PSSM. We synthesized insights about subtle details and emergent behaviors relevant to tool developers and advanced modelers. We reported inconsistencies and missing clarifications in more than 20 issues to the standardization committee. Based on these insights, we studied 11 patterns for doActivities detailing the consequences of using a doActivity in a given situation and discussing countermeasures or alternative design choices. We hope that our analysis of the semantics and the patterns help vendors develop conformant simulators or verification tools and engineers design better state machine models.
\end{abstract}

\begin{IEEEkeywords}
UML, PSSM, state machine, semantics, concurrency, pattern.
\end{IEEEkeywords}}

\maketitle

\IEEEdisplaynontitleabstractindextext

\IEEEpeerreviewmaketitle

\IEEEraisesectionheading{\section{Introduction}\label{sec:introduction}}

\IEEEPARstart{S}{tate} machines are used in engineering many types of software-intensive systems~\cite{Wagner2006, mbt-embedded, Hutchinson2011}, especially in embedded software engineering and industries like automotive or aerospace according to recent surveys~\cite{DBLP:journals/jsa/AkdurGD18, https://doi.org/10.1002/sys.21745}. There are numerous state machine variants starting from Harel's statecharts~\cite{DBLP:journals/scp/Harel87} to SCXML~\cite{barnett2015scxml}. The Unified Modeling Language (UML)~\cite{omg-uml} introduced a state machine variant specifically targeting software design, which evolved significantly in the last decades with each subsequent release. New language elements were added, and the semantics of state machines was refined, especially with the publication of the Precise Semantics of UML State Machines (PSSM) specification~\cite{omg-pssm}.

UML State Machines extend simple finite state machines with powerful constructs that help to design complex software systems. Composite and orthogonal states introduce hierarchy and concurrency; entry/exit behaviors and transition effects make it possible to describe detailed behavior. However, there is one seemingly simple and innocent language construct that fundamentally changes state machines' reactive model of computation: \emph{doActivity behaviors}.

A doActivity describes behavior that is executed \emph{independently} from the state machine once entered in a given state, typically used to model \emph{``computation or continuous activity that take time to complete and that may be interrupted by event''}~\cite{uml-reference-manual}. DoActivity behaviors are preferred~\cite{karban2018opense} to express long-running tasks that otherwise would block the event processing of the state machine. DoActivities are frequently featured in training materials for executable UML tools~\cite{magicgrid, ea-simulation}. 
Note that details of the behavior for a given state can also be modeled using submachine states or communicating state machines. However, doActivities are more convenient for expressing data-intensive computations.

\textbf{Motivation} The UML specification or textbooks are vague about how the doActivity behavior construct should be properly used. This lack of guidance is a severe issue, as a doActivity behavior is a powerful construct that can cause internal and external effects even when the state machine is in a stable configuration. When improperly used, it can introduce the worst problems: concurrent, non-deterministic bugs that are especially challenging to find and could ruin a seemingly correct software design.

\begin{figure*}[!htbp]
	\centerline{%
	\includegraphics[width=.7\linewidth]{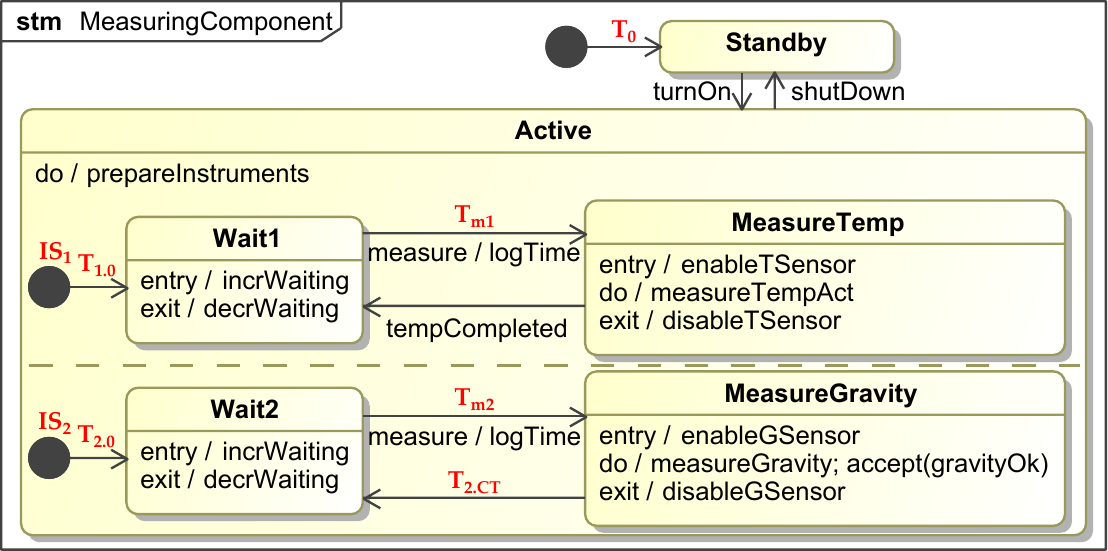}%
	\hspace{1mm}%
	\includegraphics[width=.3\linewidth]{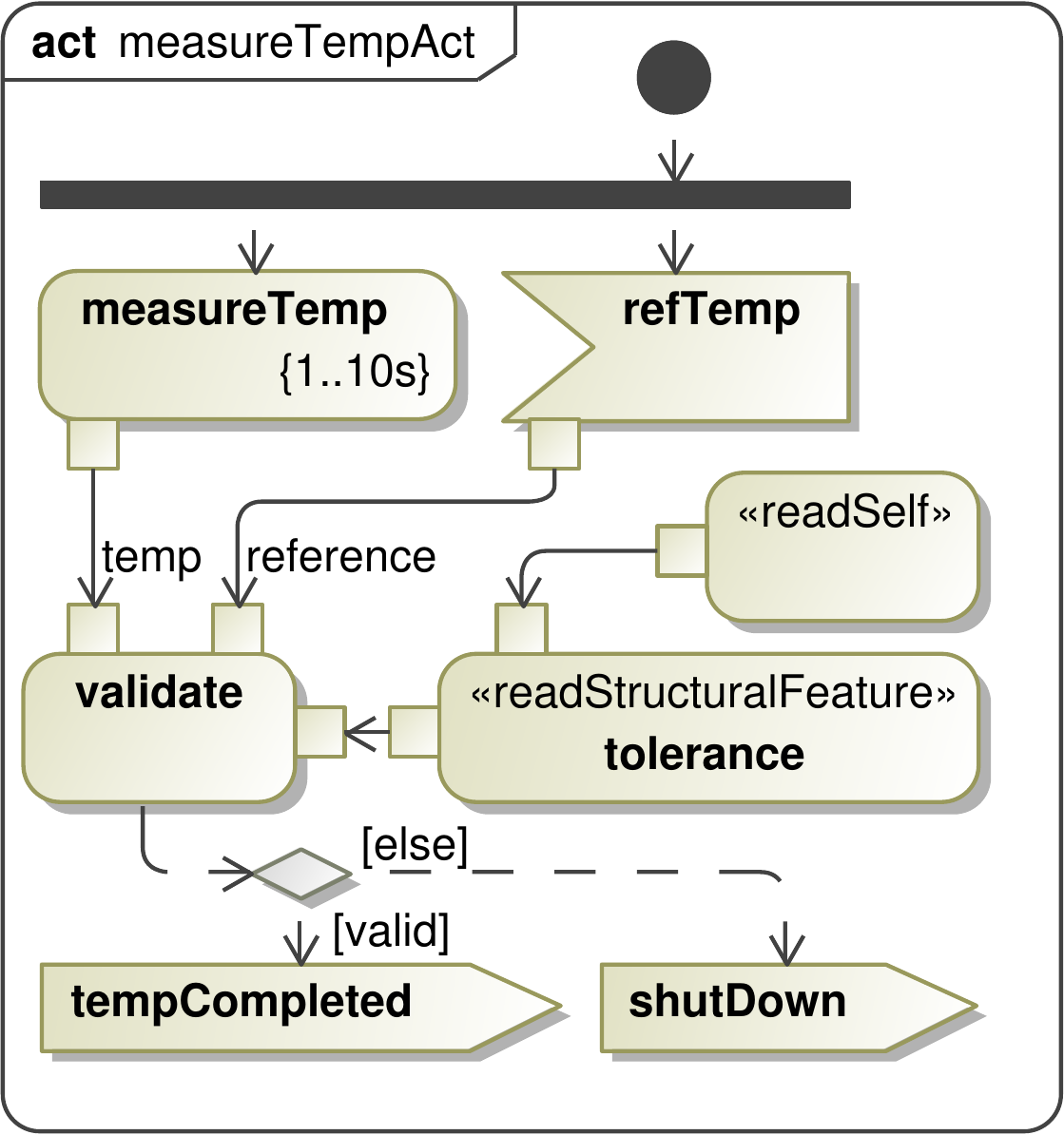}%
	}
	\caption{A state machine model of a measuring component with bad practices to show the risks of semantic misunderstandings.
	{\color{red}\bf Bold} labels show element names.
	The activity diagram details the behavior of the \elementName{measureTempAct} doActivity.
	}
	\label{fig:MeasuringComponent}
\end{figure*}

In many examples, a doActivity is just an informal note to show that some behavior is being executed in the state.
However, specifying doActivity behaviors more \emph{precisely} is especially significant if state machines are used to model detailed, \emph{executable behavior} that can be later used for simulation~\cite{DBLP:conf/sigada/Seidewitz14}, code generation~\cite{code-generation-slr} or verification~\cite{horvath-se-2023}. 
As initially UML did not have a precise way to express executable behavior, many tools used custom action languages~\cite{action-language} or extended generic programming languages~\cite{DBLP:conf/modelsward/BadreddinLF14a}.
In UML version 2.0, OMG incorporated Actions into UML as fundamental, data-flow based executable units. Several specific actions are predefined, e.g., for accessing and modifying variables or sending and receiving signals in a standardized way. Actions are defined as executable nodes of activities in the UML metamodel.
Complex sequence of Actions are expressed as activities, which can be assigned to transitions effects and state entry, exit or do behaviors in state machines. However, if some of the actions would block the event processing of the state machine, then they need to be specified in an independently running doActivity behavior.

OMG defined the Semantics of a Foundational Subset for Executable UML Models (fUML)~\cite{omg-fuml} for actions and activities, accompanied by a textual syntax (Alf)~\cite{omg-alf}. 
Building on fUML, the PSSM specification offered detailed operational semantics and execution model to answer numerous questions about the semantics of state machines. Publishing PSSM was a massive leap towards precise semantics that could be the solid basis for simulators or code generators. However, the specification is more oriented toward tool developers; ``everyday'' model users have a hard time grasping the big picture from the detailed operational rules and the subtle interactions of language elements~\cite{elekes-sqj-2023}. 

There are numerous academic works on the informal and formal semantics of UML state machines~\cite{DBLP:journals/sosym/CraneD07, 10.1145/3579821, DBLP:journals/tse/BagherzadehKJD22}. However, most of them are either for older versions (pre-PSSM) or skip the semantics of doActivities. To the best of our knowledge, there is \emph{no rigorous review} of the newest, precise semantics. Moreover, there are \emph{no clear guidelines} on how to use or not use doActivities in state machines to avoid serious errors that are later nearly impossible to catch with simulation or testing. Our goal is to provide an analysis of PSSM-based semantics targeted for advanced model users (software engineers and researchers), and patterns to help design correct state machines using doActivities.

\textbf{Method} Following our previously recommended method for assessing modeling language semantics~\cite{elekes-sqj-2023}, we specifically focused on doActivity behaviors in this paper. We collected evidence~\cite{supplementaryMaterial} by reviewing and cross-checking the text of the specification, its semantic model and executable test cases, and the simulators supporting PSSM (Eclipse Moka and Cameo Simulation Toolkit). We investigated available modeling guidelines~\cite{karban2018opense}, industrial models~\cite{url-tmt}, and repositories of open-source models~\cite{DBLP:conf/models/HebigHCRF16}. 

\textbf{Contributions} By synthesizing these sources and insights, we make the following contributions in this paper.

\begin{description}
	
	\item[Semantics] We compiled an analysis of doActivity behaviors' operational semantics from the fUML and PSSM specification highlighting previously not reported subtle details and emergent behaviors relevant to tool developers and advanced modelers. The identified challenging or ambiguous parts are backed by evidence from the normative text of the specification, PSSM's test cases or test executions in simulators (\cref{sec:deep-dive}). 
	
	\item[Patterns] Based on these insights, we systematically investigated 11 patterns for doActivities detailing the consequences of using a doActivity in a given situation, and recommending countermeasures or alternative design choices. The patterns are modular, take into account potential combinations of elements, and are built up from  simple states to state machines having doActivities in orthogonal regions (\cref{sec:do-patterns}).
	
\end{description}

The semantic insights emphasize that doActivities \emph{fundamentally change the reactive nature of state machines} by performing externally visible actions or accepting events deferred for later processing while the state machine is waiting. Moreover, as doActivities execute independently on their own thread, they introduce \emph{concurrency and non-deterministic choices} in every state machine. Therefore, engineers must always consider alternative traces and concurrency issues when adding a doActivity to a state machine. 

Our analysis and the described patterns help engineers design better state machine models, and vendors develop conformant simulators or verification tools. We reported the issues found in PSSM to OMG\footnote{\url{https://issues.omg.org/issues/spec/PSSM/1.0}}, and based on the findings, we plan to improve the future SysMLv2 language\footnote{The second author is a member of the committee responsible for the SysMLv2 specifications and co-author of~\cite{omg-kerml}.}.

Note that the UML semantics has evolved significantly in the last two decades. Moreover, many UML-based tools alter the semantics to be better suited for code generation~\cite{DBLP:journals/scp/LethbridgeFBBGA21} or verification~\cite{10.1145/3579821}, typically by reducing ambiguity and concurrency. However, in this paper we aim to present the semantics as-is in the official OMG specifications: PSSM-based state machines and fUML-based activities. We consider this relevant and useful regardless of the current practice of treating the semantics flexibly because without understanding the specification, tool conformance and interoperability can only be achieved with serious compromises.

\textbf{Structure} \cref{sec:motivating-example} illustrates state machine semantics and typical questions with doActivities. \cref{sec:deep-dive} presents a deep dive into PSSM semantics structured around the lifecycle of a doActivity. \cref{sec:do-patterns} recommends patterns for when and when not to use doActivities. The patterns are described in a practical format without going into the details of the operational semantics. Finally, \cref{sec:related-work} summarizes the related work, and \cref{sec:conclusion} concludes the paper.

\section{Illustrating state machine semantics}
\label{sec:motivating-example}

\begin{figure}[!htbp]
	\centerline{\includegraphics[width=\linewidth]{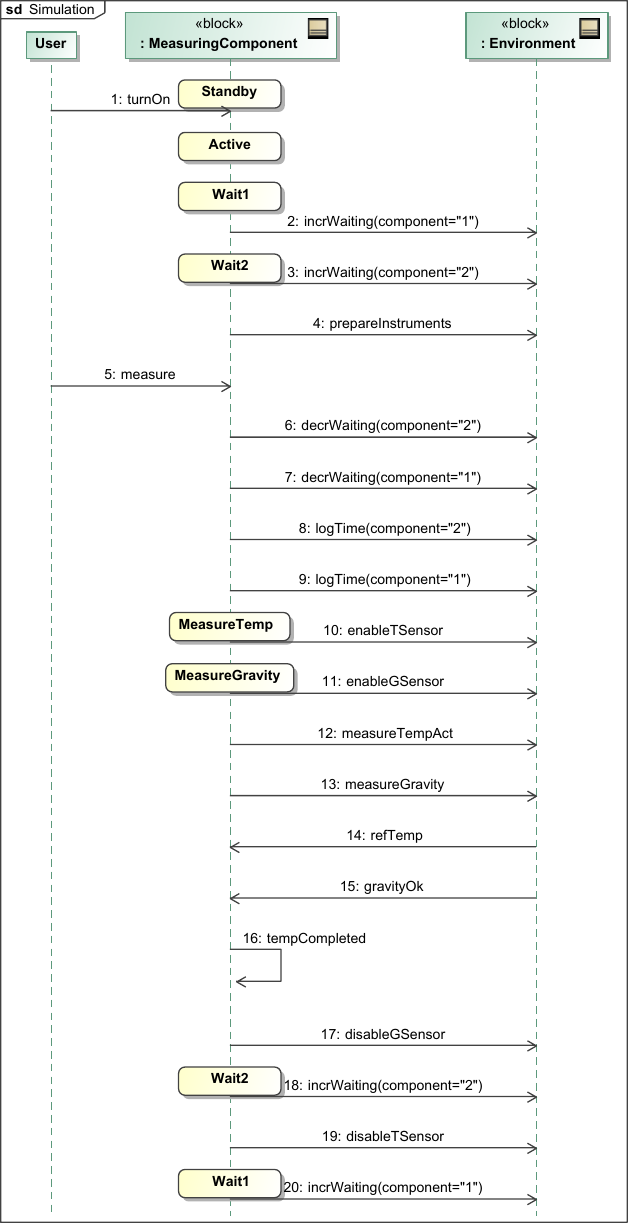}}
	\caption{Sequence diagram for simulating state machine in \cref{fig:MeasuringComponent} using Cameo.
		Input events: \event{turnOn}, \event{measure}.}
	\label{fig:MeasuringComponentSimulation}
\end{figure}

This section introduces an example state machine created for illustrating the subtle semantic questions and possible issues when using doActivity behaviors. \Cref{fig:MeasuringComponent} shows a state machine describing the behavior of a measuring component of a fictitious system inspired by the industrial modeling practices of the Thirty Meter Telescope~(TMT)~\cite{url-tmt, DBLP:conf/models/Jankevicius16}.

The component has a \stateSM{Standby} and an \stateSM{Active} state.
\emph{Orthogonal regions} model the two supported measurement types.
The component first waits for a \event{measure} signal, then performs both measurements.
The long-running behaviors are modeled in doActivities, which do the measurements and wait for external reference/confirmation signals to continue.
The temperature measurement is a complex behavior with data dependencies, therefore an activity diagram is used to describe it.
While the gravity measurement is a simple activity, defined with a textual syntax.

The example illustrates the main patterns for using doActivities in engineering practice~\cite{url-tmt, karban2018opense}: doActivity behaviors could represent long-running computations or could communicate and wait for external events that would otherwise block processing further events (i.e., placing an accept event action in an entry behavior or a transition effect could block the whole RTC step). The example is extended by outgoing log events to make the execution observable.

\newcommand{\sequenceDiagStep}[1]{\raisebox{.5pt}{\large\textcircled{\raisebox{-.2pt} {\footnotesize#1}}}}
\newcommand{\sequenceDiagWrap}[1]{#1}
\makeatletter
\makeatother

The challenging part of UML state machine semantics is that there could be many inherent \emph{nondeterministic choices} and \emph{concurrent executions}\footnote{According to UML: ``concurrent execution simply means that there is no required order in which the nodes must be executed; a conforming execution [...] may execute the nodes sequentially in either order or may execute them in parallel''~\umlPageCite{417}{15.2.3.2}} even when a simple input sequence is received. Simulation tools usually produce one execution trace for the given input: 
\Cref{fig:MeasuringComponentSimulation} shows a sequence diagram produced by a simulator tool depicting how the current state of the component changes and what signals are sent when receiving the \event{turnOn}, \event{measure} event sequence.
Let us follow this seemingly simple execution!
Circled numbers in the text denote the steps in the sequence diagram.

\begin{figure*}[htbp]
	\centerline{\includegraphics[scale=\smMicroStepsPptScale,page=1]{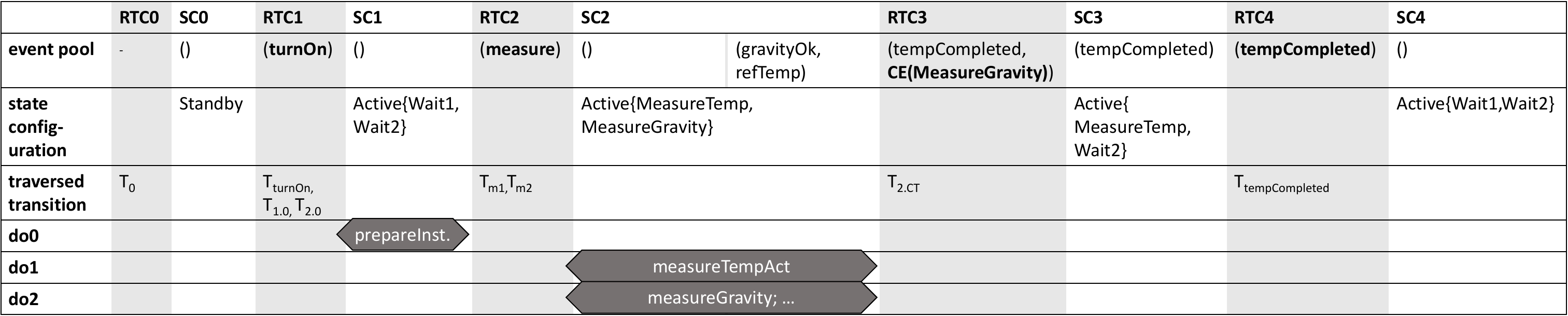}}
	\caption{Simplified execution steps based on trace in \cref{fig:MeasuringComponentSimulation}.
	Notation:
		gray columns: RTC steps,
		white columns: stable configurations, 
		boldface in event pool: dispatched event.
	(Trivial RTC steps where CEs are discarded or events are forwarded to the doActivity are omitted.)}
	\label{fig:smMicroStepsRtc}
\end{figure*}

\begin{enumerate}[label=\roman*)]
	\item The state machine alternates between stable state configurations and \emph{run-to-completion steps} (RTC steps, i.e., no other events are dispatched until the processing of the current one is completed).
	With the starting of the component begins the initial RTC step, in which firing transition \transition{T\textsubscript{0}} leaves the initial pseudostate and the component arrives at the \stateSM{Standby} state. This is the initial \emph{stable state configuration}.
	\item The state machine has an associated \emph{event pool} from which events (e.g., signal receptions) are dispatched, then matching transitions are collected and triggered.
	\item After receiving a \event{turnOn} signal \sequenceDiagStep{1} the state machine starts its first RTC step traversing several transitions forming a \emph{compound transition}~\umlPageCite{357}{14.2.3.8.4}.
	\item First, the source state (\stateSM{Standby}) is exited, then the effect of the transition is executed (not present here), and finally, the target state (\stateSM{Active}) is entered. That state is a \emph{composite state} and comprises two \emph{orthogonal regions} which are entered and executed concurrently.
	\item After the \stateSM{Active} state is entered, its doActivity behavior is started. The initial substates, \stateSM{Wait1} and \stateSM{Wait2} are entered concurrently (i.e., their order is not defined) and their entry behaviors are executed\sequenceDiagWrap{ \sequenceDiagStep{2}\sequenceDiagStep{3}}.
	At this point the RTC step initiated by \event{turnOn} is finished.
	\item The doActivity behavior \sequenceDiagStep{4} is executing concurrently with other behaviors of state \stateSM{Active}, and may continue even after the RTC step.
	\item Receiving a \event{measure} signal \sequenceDiagStep{5} starts a new RTC step and triggers firing \transition{T\textsubscript{m1}} and \transition{T\textsubscript{m2}} transitions in both regions concurrently.
	\item The \stateSM{Wait} states are exited and their exit behaviors are executed\sequenceDiagWrap{ \sequenceDiagStep{6}\sequenceDiagStep{7}}, then the transitions are traversed with their effects executed\sequenceDiagWrap{ \sequenceDiagStep{8}\sequenceDiagStep{9}}, finally the respective \stateSM{Measure} states are entered, i.e., their entry behaviors are executed \sequenceDiagStep{10}\sequenceDiagStep{11} then their doActivities are started. The RTC step completes, and a new stable configuration is reached (although the doActivities are still executing).
	\item Note that \cref{fig:MeasuringComponentSimulation} depicts one possible ordering, but there are numerous potential interleavings. In the following we describe the steps in each region separately.
	\item  The doActivity of \stateSM{MeasureTemp} performs the measurement\sequenceDiagWrap{ \sequenceDiagStep{12}}, waits for a reference value (\event{refTemp} signal\sequenceDiagWrap{ \sequenceDiagStep{14}}), successfully validates the measurement and notifies the state machine with a \event{tempCompleted}\sequenceDiagWrap{ \sequenceDiagStep{16}} signal.%
	\item The doActivity of \stateSM{MeasureGravity} is similar \sequenceDiagStep{13} but instead of self-signaling it models the behavior differently by using a \emph{completion transition}, which is triggered by the \emph{completion event} of its source state. \stateSM{MeasureGravity} state generates a completion event after all of its internal activities, i.e., here its entry and doActivity behaviors, have completed. The doActivity in this case waits until confirmation of the measurement, a \event{gravityOk} signal is received\sequenceDiagWrap{ \sequenceDiagStep{15}}, then \stateSM{MeasureGravity} is completed and it emits a completion event (not depicted). Completion events have priority over regular events.
	\item \event{tempCompleted} and the completion event trigger exiting the respective \stateSM{Measure} state and executing its exit behavior\sequenceDiagWrap{ \sequenceDiagStep{17}\sequenceDiagStep{19}}.
	In each region, the respective \stateSM{Wait} state is entered and its entry behavior is executed\sequenceDiagWrap{ \sequenceDiagStep{18}\sequenceDiagStep{20}}. %
\end{enumerate}

\Cref{fig:smMicroStepsRtc} summarizes the execution steps reconstructed based on the sequence diagram~(\cref{fig:MeasuringComponentSimulation}) in a simplified form: alternating RTC steps and stable configurations, with the actual content of the event pool and the traversed transitions. For simplicity, the figure omits those trivial RTC steps that dispatch and discard completion events in the lack of completion transitions.
This view can illustrate the  (significant) changes of the event pool, e.g. that completion events are put to the front of the pool, or that doActivities can add or handle events from the event pool even during stable configurations (see SC2). However, this figure again captures one possible execution of one simulated trace~(\cref{fig:MeasuringComponentSimulation}). We made arbitrary choices about the otherwise unobservable execution details that are not specified even in the PSSM test cases (e.g., we cannot observe the exact start or finish of the doActivity in the simulator).

This state machine seems to be unsophisticated, but even for such an example it is complicated to understand the exact order and timing of concurrent executions, or identify all valid alternative traces. Starting from the steps where doActivity and orthogonal regions are sending and receiving signals, following the executions depicted in the sequence diagram is quite convoluted. Numerous questions could arise to clarify the relations between the doActivity and the state machine, such as the following.

\begin{itemize}
	\item A doActivity starts concurrently with other behaviors in a composite state. When exactly?
	\item What happens, if a doActivity and the state machine waits for the same event? Are there priorities?
	\item A doActivity is aborted if its containing state is exited. Can it be aborted even before executing any behavior?
\end{itemize}

The UML specification describes general semantic rules, but such specific cases are sometimes hard to answer. The PSSM specification attempts to answer such questions.

\begin{figure*}[!ht]
	\centerline{\includegraphics[width=\linewidth]{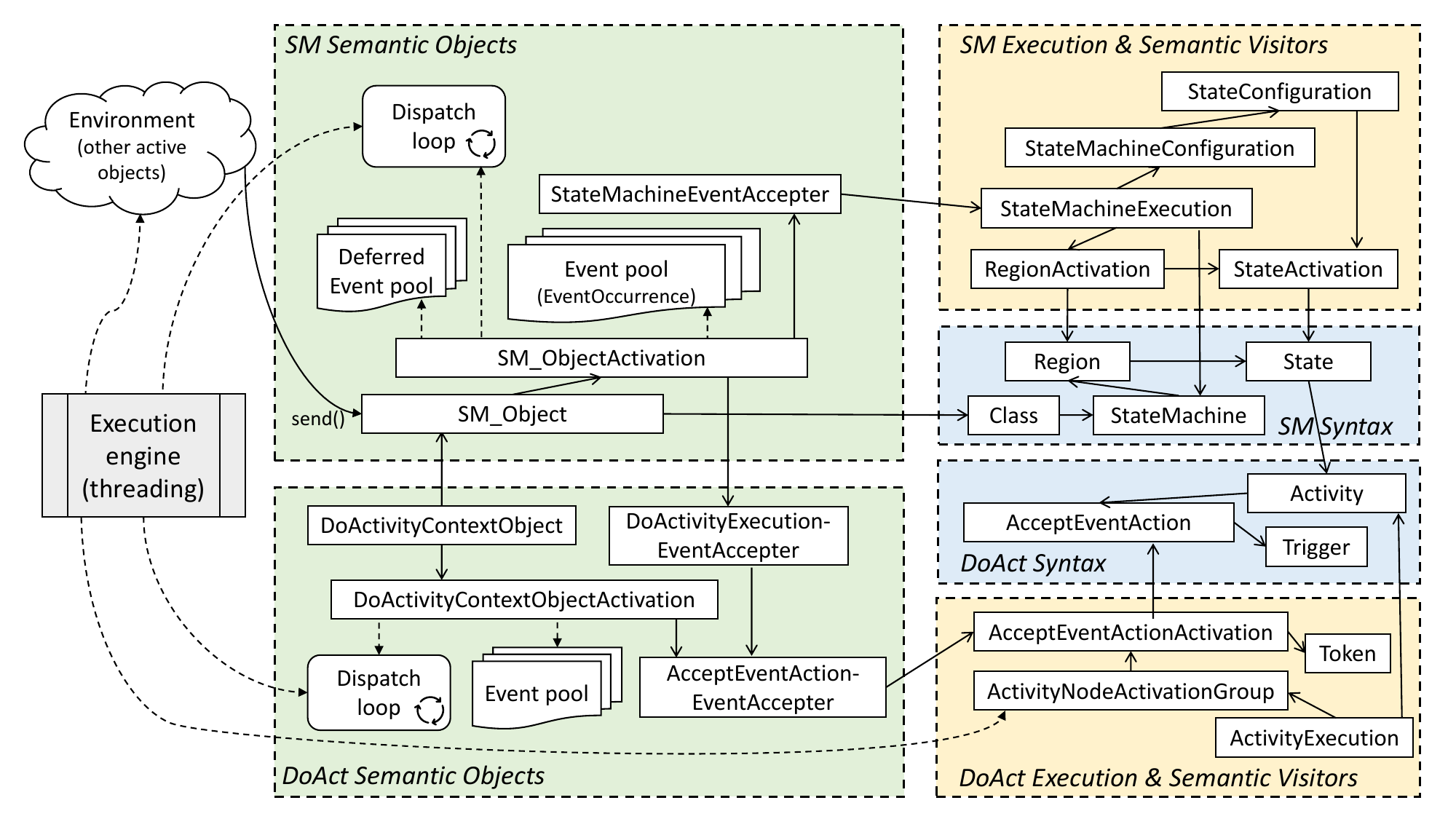}}
	\caption{Overview of the semantic classes defined in fUML and PSSM for state machines and doActivities.
	\emph{Notation:}
	rectangles: UML (meta)classes;
	other vertices: implicit concepts in PSSM;
	solid lines: associations;
	dashed lines: dependencies.}
	\label{fig:PssmSemanticClasses}
\end{figure*}

\section{Deep dive into PSSM semantics}\label{sec:deep-dive}

To understand doActivities in depth, there are two important modeling formalisms to consider: \umlClass{Activity} and \umlClass{StateMachine}, which are both \umlClass{Behavior}s.
An activity can be used to make detailed models of state machine transition and entry/exit/doActivity behaviors.
An activity defines the ordering and data dependencies of its basic steps, the \umlClass{Action}s.
UML offers a range of predefined actions, e.g., reading variables, sending signals, but arbitrarily long computations can be defined as actions as well.

This section presents the subtle details of the semantics defined in PSSM organized according to the lifecycle of a doActivity (starting, executing, event handling and finalization). Understanding doActivity semantics is especially challenging as both the event-based reactive state machine semantics from PSSM and the data-flow based activity semantics from fUML~\cite{omg-fuml} have to be considered.

Following our previously defined method~\cite{elekes-sqj-2023}, we synthesized these insights by reviewing the specification, cross-checking the definition of the operational semantics with the test cases, executing the test cases in Cameo Simulation Toolkit, and examining the source code and debugger executions of the Papyrus Moka\footnote{\url{https://marketplace.eclipse.org/content/papyrus-moka}}~\cite{guermazi:cea-01844057} reference implementation.

We highlight the parts that necessitate careful consideration when developing state machine models (e.g., concurrent behaviors or non-deterministic choices in priorities), and inconsistencies in the specification artifacts that could cause misunderstanding between engineers or tool vendors. 

The supplementary material of the paper~\cite{supplementaryMaterial} contains detailed artifacts collected from the specifications and simulator executions (e.g., tables, screenshots and models), and complex doActivities from the TMT industrial model~\cite{url-tmt}.

\subsection{Overview of the operational semantics in PSSM}
\label{sec:deepDive:overview}

The fUML specification defines an operational semantics for activities and actions, which is extended for state machines in PSSM. These specifications define an execution model for a subset of the UML language. The execution model includes an abstract execution engine, classes for event handling (\umlClass{SM\_Object} and \umlClass{EventAccepter}), and semantic visitors for each supported syntax class (e.g., \umlClass{StateActivation} for \umlClass{State}). Operations of the semantic visitors encode the semantics of the given element (e.g., \umlClass{enter} in \umlClass{StateActivation} for entering a state).

\Cref{fig:PssmSemanticClasses} captures the most important semantic classes and their connections relevant to doActivities. The following paragraphs will introduce their main role, and the subsequent subsections will go into more detail. Note that to ease understanding several classes and associations are left out (see detailed instance model in supplementary material).

\emph{Syntax}: Syntax classes are classes from the UML specification. They include classes like \umlClass{Region} and \umlClass{State} for state machines~(\emph{SM Syntax} box).
An \umlClass{Activity}~(\emph{DoAct Syntax} box) contains \umlClass{ActivityNode}s and -\umlClass{Edge}s. Nodes represent executable steps and their input/output data, while edges represent the control- and object-flow between the nodes.
An \umlClass{Action} is an executable node, the fundamental unit of executable functionality. E.g., an \umlClass{AcceptEventAction} waits for the occurrence of an event.

\emph{Execution \& Visitors}: Each syntax class has an appropriate semantic visitor (\umlClass{*Activation} classes). The visitors of the \umlClass{Activity} and \umlClass{StateMachine} top-level behavior classes are called \umlClass{ActivityExecution} and \umlClass{StateMachineExecution}, which are responsible for starting the respective behavior and serve as a container for the other visitors. The execution model includes classes for semantic concepts that are not directly represented in a UML model. For example, \umlClass{StateMachineConfiguration} describes the active state configuration for a state machine in a recursive structure, while \umlClass{Token} is an explicit representation of the token game in activities. \umlClass{ActivityNodeActivationGroup} groups nodes activated or suspended together (top-level nodes form a group, and further groups are created for structured activities).

\emph{SM Semantic Objects}: These classes represent the lifecycle and communication of the active objects. An active object is an object with a classifier behavior, while an \umlClass{SM\_Object} is an active object with a state machine as its classifier behavior. \umlClass{SM\_Object} has operations for sending/receiving event occurrences\footnote{In fUML, Event Occurrences can be asynchronous signal receptions, classifier behavior invocations (for starting behaviors), and synchronous operation calls. PSSM adds \umlClass{CompletionEventOccurrence} for completion events.} to/from other active objects in the environment or even itself. Once an instance of \umlClass{SM\_Object} is started, an \umlClass{SM\_ObjectActivation} is created for managing event handling. The superclass from fUML (\umlClass{ObjectActivation}) has an event pool: a pool of \umlClass{EventOccurrence}s that the object received. \umlClass{SM\_ObjectActivation} extends this with a second pool for deferred events~(here PSSM diverges from UML, see~\pssmPageCite{50}{p.~38}). \umlClass{EventAccepter}s can be registered to receive specific event occurrences. Dispatching events from the event pool is handled by a dispatch loop%
\footnote{The event dispatching mechanism is described in the fUML specification~\fumlPageCite{144}{8.8.1}. However, some parts of it are not specified (e.g., locking of the event pool), thus implementations can differ.}. Once an event occurrence is dispatched, the registered accepters are checked whether they match the current event. If more than one accepters match, then one of them is chosen according to a predefined choice strategy (which can be \umlClass{FirstChoiceStrategy} or even a non-deterministic choice). Contrary to activities, which register separate accepters for each \umlClass{AcceptEventAction}, \umlClass{SM\_ObjectActivation} registers \emph{only one} \umlClass{StateMachineEventAccepter} for the whole state machine, which examines the actual state configuration and handles transition priorities and conflicts.

\emph{DoActivity Semantic Objects}: PSSM introduces doActivity-specific specializations of fUML semantic classes. \umlClass{DoActivityContextObject} references the context object of the state machine (\umlClass{SM\_Object}) and can access its structural features. As event occurrences cannot be sent directly to the doActivity~\pssmPageCite{52}{8.5.6}, handling events is a two-phase process. If the doActivity wants to wait for an event, it first registers typically an \umlClass{AcceptEventActionEventAccepter} to its \umlClass{DoActivityContextObjectActivation}. Next, it encapsulates the original accepter in a \umlClass{DoActivityExecutionEventAccepter}, and registers the encapsulating accepter with the state machine. If the state machine dispatches an event that matches the doActivity's accepter, then the encapsulating accepter adds the event occurrence to the event pool of the doActivity. Once that event occurrence is dispatched asynchronously in the context of the doActivity, the original accepter can handle it in an RTC step of the doActivity~\fumlPageCite{144}{8.8.1}.

\begin{figure*}[htbp]
	\centerline{\includegraphics[scale=\smMicroStepsPptScale,page=2]{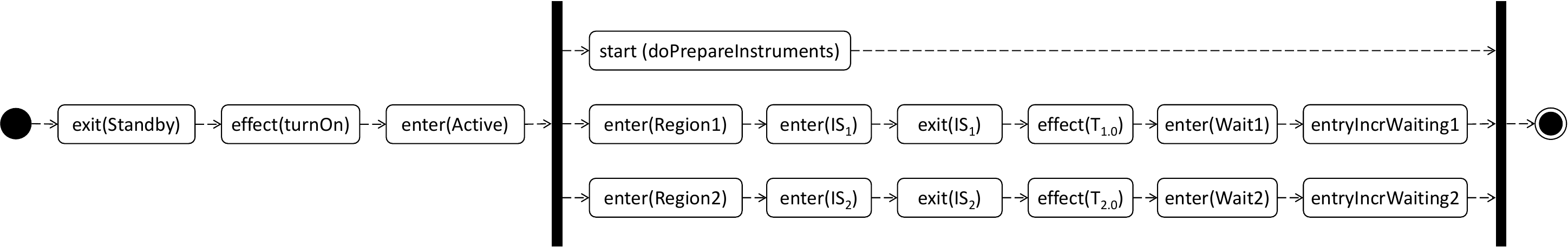}}
	\caption{Ordering constraints of concurrent steps during RTC1 from \cref{fig:MeasuringComponentSimulation}. The doActivity is only started, but its execution takes place asynchronously on its own thread. Note that these actions are not atomic, their internal steps can interleave.}
	\label{fig:smMicroStepsOrthogonalRegionConstraints}
\end{figure*}

\emph{Execution model \& threading}:
The fUML and PSSM specifications define a very generic, concurrent execution model, which leaves room for diverse implementations but hinders the understanding of how models can behave and how execution tools should be implemented.
Tools have to schedule the overlapping execution of each active object.
Though the specifications lack an explicit definition of execution threads, the tools need to define some kind of ``threads'' and orchestrate them.
The thread concept is rather generic, and its realizations can vary, ranging from sequential programs or OS-threads to running threads on separate physical processors or even separate servers~\umlPageCite{483}{p. 441}.
Most of the execution model is defined using partial ordering constraints, e.g., an RTC step is not finished until the entry behaviors are executed.
The specification explicitly states that active objects, doActivities, and transmissions of each EventOccurrence run asynchronously on their own threads.
These executions can be fully parallel, i.e., the actions of a doActivity can be executed in parallel with the actions of the state machine's transition effect\footnote{As there is no parallel implementation available, we could not cross-check this. For example, Moka defines meta-tasks for event sending and event dispatching, and schedules these tasks sequentially. However, the traces of PSSM tests Transition 019 and Entering 011 suggest such concurrency.}.
Actions not contained in isolated regions can see intermediate results of other activities~\fumlPageCite{232}{8.10.1}, and even simple actions might not be atomic~\umlPageCite{485}{16.2.3.1}.
For example, to realize the example state machine~(\cref{fig:MeasuringComponent}) in a fully parallel way, we need separate execution threads for the state machine's event accepter, both orthogonal regions, all running doActivities, and even all concurrent actions in the activity.

The plethora of these concurrency problems might not be evident for UML users because execution tools do not have to warn that a step is not atomic and might be interrupted, or recognize and execute all possible concurrent executions in parallel; as long as a tool produces a legal trace satisfying the partial orderings, the tool conforms to the specification. Thus various simulators or code generators might produce surprisingly different traces for the same state machine.

\vspace{6pt}
\noindent\fbox{%
	\parbox{0.97\linewidth}{%

		\textbf{Insights} Threading in fUML/PSSM is permissive:
		\begin{itemize}
			\item A simple state machine has one execution thread, triggered by dispatching a new event.
			\item However, a transition effect, entry/exit behavior defined by an activity could contain concurrent actions.
			\item Moreover, if there is an orthogonal region, then behaviors inside the different regions are concurrent and can interleave with each other (even the exit--effect--entry steps in a transition firing are not atomic).
			\item Each doActivity has its own execution thread, independent from the RTC step of its state machine.
		\end{itemize}
}}

\subsection{Starting a doActivity behavior}
\label{sec:deepDive:starting}

A doActivity commences execution when the State is entered and the \emph{entry} Behavior has completed~\cite[14.2.3.4.3]{omg-uml}. The important details are the following:

\begin{itemize}
	\item what exactly ``starting a doActivity'' means,
	\item what steps are finished in the state machine's RTC step,
	\item what behaviors are concurrent with the doActivity.
\end{itemize}

The UML specification explicitly states that the execution of a doActivity is concurrent with the entry Behaviors of substates. The fUML and PSSM specification refines and clarifies what ``commencing execution'' means. When a classifier behavior is created and started, then its execution does not run immediately. Instead, during \elementName{startBehavior} the object activation of the doActivity registers a specific \umlClass{ClassifierBehaviorInvocationEventAccepter} and adds an \umlClass{InvocationEventOccurrence} to its event pool\footnote{This mechanism was added in fUML 1.2, and its rationale is explained in the issue \href{https://issues.omg.org/issues/FUML12-35}{FUML12-35}.}. These steps are performed in the RTC step of the state machine. Later, this invocation event occurrence is dispatched asynchronously, and handled in a so called initial RTC step of the doActivity, where elements of the activity are activated and fired.

\begin{figure}[htbp]
	\centerline{\includegraphics[scale=\smMicroStepsPptScale,page=3]{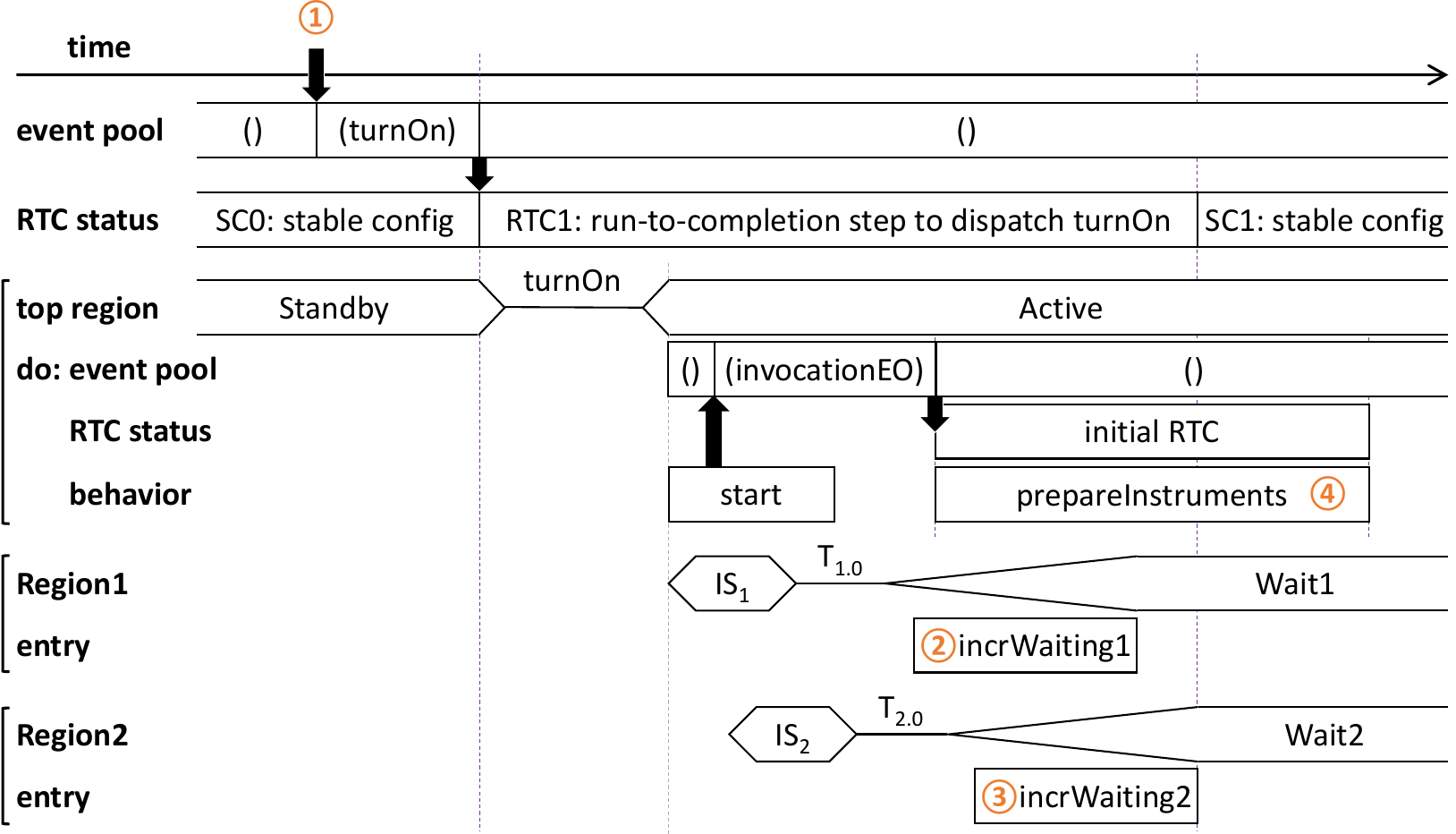}}
	\caption{Detailed steps for entering the \stateSM{Active} state. The \elementName{prepareInstruments} doActivity is started, then its behavior is asynchronously executed. Circled numbers in orange represent events observed during simulation~(\cref{fig:MeasuringComponentSimulation}).} 
	\label{fig:smMicroStepsTimeline}
\end{figure}

\Cref{fig:smMicroStepsTimeline} illustrates starting a doActivity by showing a timeline of a possible order of state changes and execution steps consistent with the sequence diagram observed during simulation~(\cref{fig:MeasuringComponentSimulation}).
The figure shows the event pool, active states in each region, and their associated behaviors being executed. The example points out two challenges when reasoning about even one possible trace of a state machine.

\begin{enumerate}
	\item \emph{Lack of standard notation}: We used a format similar to timing diagrams, but even that needed to be modified, e.g., to depict that entering a state takes time while the entry is executed. Having such a detailed view is essential to understand the relations of each step.
	\item \emph{Observability issues}: The figure features several arbitrary choices that cannot be derived from simulation results~\cite{elekes-sqj-2023}, e.g., the exact start and end of steps in the activities, especially when the doActivity finishes.
\end{enumerate} 

\Cref{fig:smMicroStepsTimeline} only depicts one possible ordering for a given trace, but due to concurrency, numerous steps could interleave. We recommend thinking about partial orders and alternate traces for an RTC step.  \Cref{fig:smMicroStepsOrthogonalRegionConstraints} focuses on the ordering constraints of concurrent steps executed during RTC1: even starting the doActivity can be in parallel with entering each of the sub-regions.

\vspace{6pt}
\noindent\fbox{%
	\parbox{0.97\linewidth}{%
		\noindent \textbf{Insights} Nothing is guaranteed to be executed from the doActivity during the state machine's RTC step, and even starting the doActivity is concurrent with behaviors from substates. There is no guarantee that anything will ever be executed from the started doActivity (see \cref{sec:deepDive:finalization}).
}}

\subsection{Executing a doActivity behavior}
\label{sec:deepDive:executing}

As discussed in \cref{sec:deepDive:starting,sec:deepDive:overview}, a doActivity executes asynchronously from the state machine. This execution can happen concurrently with state machine RTC steps (the one in which it was started and possibly later steps as well) or even between RTC steps when the state machine is waiting. 

If executed during a state machine RTC step, then the doActivity can interleave\footnote{Potential interleaving is exemplified by PSSM test cases Transition 017 or Deferred 006-C. Note that any part of the doActivity can interleave, and not just parts that are after waiting for an event, as some PSSM tests might suggest.} with any other behaviors that may be associated with the State~\umlPageCite{351}{14.2.3.4}. Specifically, doActivity runs concurrently with
\begin{itemize}
	\item other doActivities in any active states (either parent states, nested substates, or other orthogonal regions),
	\item entry/exit behaviors and transition effects in other regions or nested substates,
	\item internal transitions in the doActivity's state, other regions, parent states or nested substates.
\end{itemize}

Moreover, the doActivity can send signals to external components during execution. If this happens between RTC steps, then the state machine performs externally visible actions in what should be stable configurations.

\vspace{6pt}
\noindent\fbox{%
	\parbox{0.97\linewidth}{%
		\textbf{Insights} DoActivity runs during and between RTC steps.
		\begin{itemize}
			\item Parts of a doActivity can interleave with any running behaviors of the state machine or other doActivities.
			\item Externally visible actions in a doActivity can continue execution after the RTC step of the state machine.
		\end{itemize}		
}}

\subsection{Handling events in a doActivity behavior}
\label{sec:deepDive:events}

Since doActivities can accept and wait for events, the exact timing, ordering and priorities are important to understand.

\textbf{Timing of accepters}
A state machine registers a single event accepter as soon as it starts. A doActivity registers an event accepter only after it fires an accept event action during its execution.
The fact that the doActivity can only receive events from the state machine's event pool alters its behavior compared to a standalone activity.
1) The doActivity may receive an event that was pending in the state machine's event pool even before the start of the doActivity. 
2) If there are long-running actions before the accept action, then a doActivity can ``miss'' an event it could handle because the state machine may discard the event due to the lack of a matching accepter.

\textbf{Conflict and priorities}  Without doActivities, there is only a single \umlClass{StateMachineEventAccepter} responsible for handling events. 
Using doActivities waiting for events changes this simple situation: doActivities can register competing event accepters; doActivities in composite states can each register their accepters at the same time, and even one doActivity can register several accepters if it contains multiple concurrent \umlClass{AcceptEventAction}s.
 
These multiple accepters can wait for the same event, resulting in a \emph{conflict}. One might expect that there is a clear conflict resolution strategy similar to conflicting transitions and substate's behavior always overrides the parent state's behavior. However, from the semantics of the \umlClass{dispatchNextEvent} operation of \umlClass{ObjectActivation} it seems to us that there are \emph{no priorities} between these accepters and one accepter is chosen \emph{non-deterministically}\footnote{More precisely, the choice is made according to the current choice strategy. Even if \umlClass{FirstChoiceStrategy} is used, the result is hard to predict, as the order in which concurrent doActivities register accepters is arbitrary. Moreover, the state machine's accepter gets removed and re-registered after each event dispatch, thus it can be the very last registered accepter in some cases. See supplementary Moka screenshots~\cite{supplementaryMaterial}}. There are no explicit PSSM test cases exemplifying this but simulation in Moka~\cite{supplementaryMaterial} confirms competing accepters%
\footnote{Test Behavior~003-B shows that the doActivity can accept an event due to the lack of transitions triggered by the event. Test Deferred 006\=/C illustrates that if two doActivities  wait for the same event, then one of them is selected non-deterministically (%
unlike transitions, which can use the same event to traverse transitions in different regions). However, there are no tests where a doActivity and the state machine compete for the same event (Deferred~006-B contains such a case, but the text erroneously states that there are no transitions enabled).}.

\textbf{Defer} Some conflicts can be resolved by deferring as suggested by the specification\footnote{``\ldots in general, an executing doActivity Behavior will compete with the executing StateMachine that invoked it to accept EventOccurrences dispatched from the same eventPool. Nevertheless, in some situations it is necessary to ensure that a doActivity is able to accept certain EventOccurrences instead of the StateMachine. To allow this, a deferredTrigger should be used on the State that owns the doActivity, in which case any EventOccurrences deferred while the StateMachine is in that State may be consumed by the executing doActivity.'' \pssmPageCite{52}{8.5.6}}. The \umlClass{defer} keyword can be used to ``capture'' an event in a given state, place it in the deferred pool, and only release it once the state deferring the event is left.
A state machine can defer an event if
\begin{itemize}
	\item an active state defers the event,
	\item there is no enabled transition with a \emph{higher priority} from the active state configuration, and %
	\item there is no enabled \emph{``overriding''} transition, i.e., a transition from the deferring state.~\pssmPageCite{65}{8.5.9}	
\end{itemize}

Priorities between transitions and defer presumably follow the same principles as firing priorities between transitions: transitions from direct and indirect substates have priority over transitions from containing states, while other transitions have no priority difference~\umlPageCite{359}{14.2.3.9.4}. 

In the presence of doActivities the above is extended with two rules~\pssmPageCite{52}{8.5.6}.

\vspace{6pt}
\noindent
\emph{1) If the state machine is about to defer an event for which a doActivity has also registered an accepter, the state machine is not allowed to defer it, and the event can be accepted by the doActivity.}

\begin{figure}[htbp]
	\centerline{\includegraphics[scale=\plantumlscale]{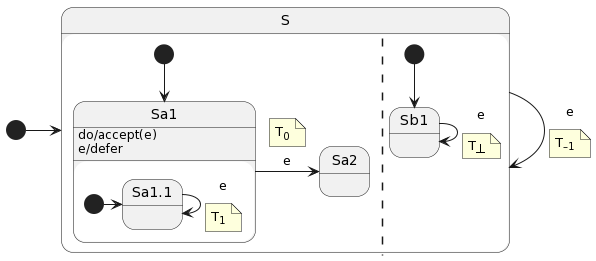}}
	\caption{DoActivity and transitions triggered by the same event~\event{e} with event deferral in state \stateSM{Sa1}.}
	\label{fig:accept-vs-transition-with-defer}
\end{figure}

To analyze priorities when using defer, \Cref{fig:accept-vs-transition-with-defer} shows transitions all triggered by event \event{e} in various positions to the deferring state \stateSM{Sa1}.
Consider that the state machine is in \stateSM{S}[\stateSM{Sa1}[\stateSM{Sa1.1}],~\stateSM{Sb1}] state configuration. Assume that each transition marked with a note exists in a separate state machine on its own (so that they are only in conflict with the defer and not with each other).
There are two cases: 

\begin{enumerate}[label=\Alph*)]
	\item \transition{T\textsubscript{--1}} has lower priority, and \transition{T$_\perp$} in an orthogonal region (other than the deferring state) has no order of priority.
	Thus the state machine is about to defer but is not allowed due to the waiting doActivity accepter. The result is that the state machine's event accepter does not match the event and the doActivity can accept it.
	
	\item \transition{T\textsubscript{0}} is an overriding transition, and \transition{T\textsubscript{1}} originating from a nested state has higher priority.
	Therefore, no defer occurs and the state machine matches the event.
	The result is that both the state machine and the doActivity match the event and therefore -- possibly against the modeler's intention -- one of the accepters is chosen non-deterministically.
	In fact, the presence of an unguarded \transition{T\textsubscript{0}} renders the defer completely useless.
	With a guarded {T\textsubscript{0}} or any transition like {T\textsubscript{1}}, the defer will be ineffective whenever these transitions are enabled. 
\end{enumerate}

To summarize, \umlClass{defer} will give priority to the doActivity only in case A)\footnote{We were not able to cross-check this conclusion using PSSM tests, as there is no such specific example. Closest tests are \pssmPageLink{214}{Deferred~002} and \pssmPageLink{215}{Deferred~003} but without doActivities.}. Note, however, that there is no mechanism to give priority to the state machine's transitions.

\vspace{6pt}
\noindent
\emph{2) If the state machine has already deferred an event for which the doActivity registers an accepter, then that event can be accepted by the doActivity directly from the deferred event pool~\pssmPageCite{52}{8.5.6}.}
\vspace{3pt}

However, it is not discussed in PSSM when this event acceptance happens with respect to the state machine (during an RTC step or between RTC steps in stable configurations). On the one hand, both the semantic description of PSSM and the code of Moka confirm that event accepters in doActivity can remove events from the deferred pool directly when registering without considering the state machine's status. As a doActivity has an independent thread of execution, removing an event from the deferred pool can happen between RTC steps without any reaction in the state machine (see test \pssmPageLink{225}{Deferred~006-B}, where there is no respective RTC step for the state machine). On the other hand, the state machine can perform an RTC step when the doActivity is about to register an accepter for an event in the deferred pool. Is the doActivity allowed to remove the event in this case? It is hard to have a definite answer as the locking mechanism for event pools is not described in the standards. However, Moka for example, implements the deferred pool as a simple \umlClass{List} without any mutual exclusion.

\vspace{6pt}
\noindent\fbox{%
	\parbox{0.97\linewidth}{%
		\textbf{Insights}: Event handling in doActivities is complicated.

		\begin{itemize}
			\item DoActivities register accepters during execution and not when starting; thus, some events can be ``missed''.
			\item Event accepters for the state machine and doActivities \emph{always} compete for events. In conflicting situations, a non-deterministic choice is made unless defer is used (but even defer resolves only some cases).
			\item There is no mechanism to give priority to the state machine's transitions over the doActivity.
			\item DoActivity can remove events from the deferred pool even during RTC steps and between RTC steps.
		\end{itemize}
}}

\subsection{Finalization of a doActivity behavior}
\label{sec:deepDive:finalization}

The execution of a doActivity can be finalized in two ways:

\begin{itemize}
	\item \emph{Completion}: the doActivity completes its execution,
	\item \emph{Destruction}: its containing state is exited, and the doActivity is aborted.
\end{itemize}

In the completion case, the execution finishes naturally (completing all node activations or executing a final node), and a completion event may be generated. Note that PSSM contains a specific clause that after a doActivity RTC step, if there are no more event accepters registered, then the doActivity completes and notifies its \umlClass{SM\_Object}~\pssmPageCite{52}{8.5.6}.

Regarding the destruction case, as the doActivity starts its execution asynchronously, the doActivity may not even execute any of its actions before aborting. This is illustrated in PSSM test cases \pssmPageLink{125}{Event~017\=/B} and \pssmPageLink{192}{Terminate~002}~\footnote{Note that the Behavior 003-A test case shows that the doActivity is only aborted after waiting for an event, but we think it is an error and the test case misses an alternative trace.}. 

Moreover, as UML explicitly states that actions are not atomic, it would be important to see whether an abort can happen during the execution of an action (e.g. for long-running actions or for modifying structural features). However, there are no such PSSM test cases, and it is not possible to validate this hypothesis with currently available non-parallel implementations.

The doActivity is always aborted before executing the exit behavior of its state. The \umlClass{destroy} call responsible for abortion is likely a synchronous call (``Exiting a StateActivation involves the following sequential steps''~\pssmPageCite{49}{8.5.5}). But properly implementing this synchronization is not trivial.

\vspace{6pt}
\noindent\fbox{%
	\parbox{0.97\linewidth}{%
		\textbf{Insights}: A doActivity can be destroyed any time during its execution, possibly even during an action.

		\begin{itemize}
			\item DoActivity can be aborted before executing anything.
			\item DoActivity must be fully aborted before the exit behavior is executed, which needs synchronization.
		\end{itemize}
}}

\subsection{Takeaway messages}

As this section explained in detail, having a doActivity in a state machine has fundamental consequences on the computation model applied, which is usually not evident for engineers designing and modeling with state machines.

Taking into account these insights about the semantics, we might wonder whether there are any issues with the state machine of \cref{fig:MeasuringComponent}. For example, once we know that a doActivity can be abruptly aborted, can we be sure that \elementName{prepareInstruments} finishes preparations before a measurement? How should we refactor it? Would it be better to place this activity in the entry behavior?\footnote{If \elementName{prepareInstruments} contains an accept waiting for an event, then placing it in the entry is definitely a bad practice. The entry behavior is executed as part of the state machine's RTC step, which blocks dispatching further events. Thus it will cause a deadlock, according to our understanding. However, Cameo Simulation Toolkit waits for and dispatches a new event even in an entry, while Moka skips the \umlClass{AcceptEventAction} and completes the activity.} The next section answers such questions by proposing practical patterns for using doActivities in certain modeling situations.

\vspace{6pt}
\noindent\fbox{%
	\parbox{0.97\linewidth}{%
		\textbf{Key insights} Using doActivities fundamentally changes the reactive nature of state machines.
		
		\begin{enumerate}
			\item doActivities can compete with the state machine for events (without doActivities, there is always only one event accepter for the state machine).
			\item doActivities can perform externally visible actions between the state machine's RTC steps.
			\item doActivities can accept previously received and deferred events without initiating an RTC step. 
		\end{enumerate}
}}
\vspace{6pt}

\section{DoActivity patterns}\label{sec:do-patterns}

We systematically collected patterns describing how doActivities can -- but not necessarily should -- be used in state machines w.r.t other modeling elements (e.g., from using a doActivity in a simple state to multiple doActivities in regions).
In order to cover important combinations that could appear in user models, the patterns are derived from a syntax point of view.
However, they are not meant to be design patterns, more like patterns on a checklist to be used in model reviews or automated static analysis tools.

For each pattern, we list possible intentions why a modeler could use a doActivity in the given context. Based on insights from previous sections, we highlight consequences and issues, then discuss potential countermeasures and advice.
The patterns are modular, and only the more significant issues are repeated. Therefore, advice should be combined if more patterns are applicable.

\subsection{Method of designing the patterns}
\label{sec:do-patterns:Method}

\begin{figure*}[htbp]
  \centerline{%
  \resizebox{0.9734\linewidth}{!}{%
    \footnotesize%
    \input{pattern-design-crop.tex}}}
  \caption{
    Classification tree~\cite{DBLP:journals/stvr/GrochtmannG93} for the patterns combining doActivities and other modeling elements.
    \emph{Notation:}
      \CIRCLE{} modeling element is present in the pattern,
      \Circle{} only the text covers the modeling element.
    $\ast$ The speciality of \labelcref{sec:guideline-DoInSimple-SelfSignaling} is self-signaling using an outgoing event. $\ast\ast$ The speciality of \labelcref{sec:guideline-DoInSimple-InternalTransition} is that the doActivity can affect the execution of an internal transition.
  }
  \label{fig:pattern-design}
\end{figure*}

To cover the different behaviors a doActivity can exhibit, we collected the modeling elements that can affect the behavior of a doActivity.
\cref{fig:pattern-design} shows a classification tree, which is used in software testing to classify relevant aspects and combine them into test cases~\cite{DBLP:journals/stvr/GrochtmannG93}.
The tree lists the aspects and partitions them into complete, disjoint choices, e.g., whether the doActivity is located in a simple or a composite state. Some choices are further refined, e.g., whether the composite state has a single or more regions and where a doActivity is located, in a parent/substate or both.
The rows represent the patterns in this section and what choices they combine from the modeling elements used for classification.

The patterns cover all single choices (all columns) for modeling elements.
For example, the patterns systematically cover doActivities in all possible locations, i.e., in a simple state~(%
{%
\renewcommand{\crefpairconjunction}{,~}%
\cref{sec:guidelineCategory-DoInSimple,sec:guidelineCategory-DoAccepting}%
}%
), in a parent/substate and both of a composite state with a single region~(\cref{sec:guidelineCategory-DoInComposite}), and in both parent and substates of orthogonal regions~(\cref{sec:guidelineCategory-DoInCompositeAndOrthogonal}).
As the number of all possible combinations of the modeling elements would be too high, we focused on first covering each leaf choices in isolation in simple states. Later in patterns with composite states we combined multiple aspects.
The modularity of the patterns allows combining them to cover the remaining combinations when more patterns match the state machine under analysis.
Empty circles denote the elements that are not present in the pattern but the description refers to previous patterns already covering that case.

\subsection{DoActivity in a simple state}
\label{sec:guidelineCategory-DoInSimple}

\guidelineSect{Basic case}
\label{sec:guideline-DoInSimple-NoAccept}

\begin{figure}[htbp]
  \centerline{\includegraphics[scale=\plantumlscale]{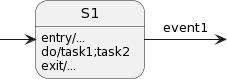}}
  \caption{A state with doActivity and transition.}
  \label{fig:DoInSimple-NoAccept}
\end{figure}

\guidelineSubsect{Pattern}
In~\cref{fig:DoInSimple-NoAccept} a simple state has a doActivity behavior (without an accept event action). The state has an outgoing transition (external transition with explicit trigger).

\guidelineSubsect{Possible intention}
In the state there is a long-running activity representing an idle task or an interruptable, non-blocking background process, which can run while in \stateSM{S1}.

\guidelineSubsect{Potential issues}
It is important to note that this pattern means the implication ``while in \stateSM{S1}, the doActivity can run'' and not the reverse ``while the doActivity runs, \stateSM{S1} is active''. 
In fact, the doActivity is an independent unit of execution with its own lifecycle.

\begin{warnEnum}
  \item\label{warn:lateStart}
  There is no guarantee when it will actually perform its actions. We only know that it starts the execution after the state's entry behavior has completed (if any).
  \item\label{warn:abortEvenBeforeRun}
  The doActivity is aborted once the state is exiting and we cannot be sure what part of the execution, i.e., which actions have finished before the abort. In the worst case, the doActivity might have not performed anything.
  \item\label{warn:abortDuringAction}
  There is no guarantee about the atomicity of actions in doActivities, i.e., it is possible that the doActivity is aborted during \elementName{task1} and only parts of it have finished.
\end{warnEnum}

\guidelineSubsect{Discussion}
If the intention is to ``wrap'' the execution of the doActivity in a state, or the intention was to wait for the doActivity to finish and then wait for the arrival of an \event{event1}, the model should explicitly enforce this by either using a \emph{completion transition}~(\cref{sec:guideline-DoInSimple-NoAccept-Completion}) or \emph{self-signaling}~(\cref{sec:guideline-DoInSimple-SelfSignaling}).

Also note that how abort is implemented is not a precisely defined part in UML.
During development of execution environments, special care should be taken to enforce the rules for abort, e.g., the execution of the doActivity is fully aborted before running the exit behavior.

\guidelineSect{DoActivity and completion transition}
\label{sec:guideline-DoInSimple-NoAccept-Completion}

\begin{figure}[htbp]
  \centering
  \subfloat[Basic case]{\includegraphics[scale=\plantumlscale]{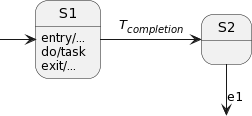}%
    \label{fig:DoInSimple-NoAccept-Completion}}
  \qquad
  \subfloat[Abort and deferred events]{\includegraphics[scale=\plantumlscale]{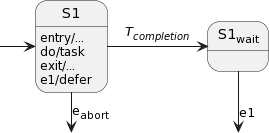}%
    \label{fig:DoInSimple-NoAccept-Completion-Defer}}
  \caption{A doActivity and a completion transition.}
\end{figure}

\guidelineSubsect{Pattern}
In~\cref{fig:DoInSimple-NoAccept-Completion} a simple state has a doActivity behavior (without an accept event action). The state has an outgoing completion transition to move to the next state~(\stateSM{S2}) once the doActivity has finished. \stateSM{S2} waits for event \event{e1} to continue.

\guidelineSubsect{Possible intention}
In the state there is a long-running activity that the modeler wants to be completely finished before continuing on event \event{e1}.
Thus, a single completion transition (an outgoing transition without a trigger) is used.

\guidelineSubsect{Potential issues}
A single completion transition fulfills the expectation that the doActivity should be finished before moving to the next state~(\labelcref{warn:lateStart,warn:abortEvenBeforeRun,warn:abortDuringAction}).

\begin{warnEnum}
  \item\label{warn:doActivityForever} In the lack of other outgoing transitions from~\stateSM{S1} the state machine will stay indefinitely in~\stateSM{S1} if the doActivity runs forever. There is no means to abort.
  \item\label{warn:loseEventWhileInDo} If there are incoming events, the state machine will dispatch and discard them due to the lack of an enabled transition. This means an \event{e1} arriving before an unusually long doActivity completes will be missed.
\end{warnEnum}

\guidelineSubsect{Discussion}
\cref{fig:DoInSimple-NoAccept-Completion-Defer} shows a possible solution for the issues above.
Explicit outgoing transition(s)~(\event{e\textsubscript{abort}}) can be used to abort the doActivity and exit state \stateSM{S1}.
Defer is used to collect events~(\event{e1}) that need to be processed after the doActivity finishes~(see \cref{sec:guideline-DoAccepting-Defer}).

Based on gray literature\footnote{For example: Completion transitions and implicit triggers -- \url{https://www.webel.com.au/node/2651}} and our experience, some engineers suggest avoiding completion transitions, partly because a transition without a trigger might be confusing for users familiar with other state machine variants\footnote{SCXML uses \texttt{done.state.id} for similar constructs.}. Self-signaling might be an alternative solution~(\cref{sec:guideline-DoInSimple-SelfSignaling}).

\guidelineSect{DoActivity using self-signaling}
\label{sec:guideline-DoInSimple-SelfSignaling}

\begin{figure}[htbp]
  \centerline{\includegraphics[scale=\plantumlscale]{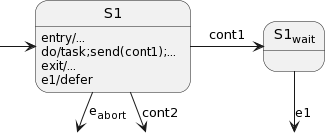}}
  \caption{A doActivity with self-signaling.}
  \label{fig:DoInSimple-NoAccept-SelfSignaling}
\end{figure}

\guidelineSubsect{Pattern}
In~\cref{fig:DoInSimple-NoAccept-SelfSignaling} a simple state has a doActivity behavior (without an accept event action). The doActivity notifies the state machine about the completion by sending a signal. The state uses \event{cont} events to continue to the next states.
Defer is used to collect events to be processed after the doActivity.

\guidelineSubsect{Possible intention}
The doActivity can signal the completion of its behavior to the state machine by sending signals. Outgoing transitions triggered by these signals are used to leave the state. Self-signals are similar to a completion event but here the doActivity can use multiple signals~(\event{cont*}) to differentiate based on results of its execution.

\guidelineSubsect{Potential issues}
Lack of abort and missed events are solved similarly as in~\cref{sec:guideline-DoInSimple-NoAccept-Completion}. However, self-signaling introduces other problems.

\begin{warnEnum}
  \item\label{warn:SelfSignal-ReorderDelayLoseDuplicate}
  The \event{cont*} events will be put into the event pool like any other incoming signal and will be processed later, once they get dispatched.
  As signals are sent asynchronously, they can be arbitrarily delayed or reordered~\fumlPageCite{145}{8.8.1}.
  Moreover, the fUML/PSSM execution model has no assumption that the communication is reliable and the signals are never lost or duplicated~\fumlPageCite{18}{2.3}.
\end{warnEnum}

\guidelineSubsect{Discussion}
Self-signaling has the advantage that all transitions have explicit, named triggers. This could be important in case of complex doActivities, which can send different signals to trigger different transitions~(\event{cont*})~\cite{OpenSE-CookbookV2}.

Using a completion transition -- if available in the modeling tool -- solves~\labelcref{warn:SelfSignal-ReorderDelayLoseDuplicate}. Otherwise, such signals should be handled specially to ensure that they are not delayed or lost.
In addition to defining internal modeling guidelines, the whole toolchain should guarantee the special behavior of the self-signaling event, e.g., simulators, code generators, test environments or manual implementations.

\guidelineSect{A state with internal transition}
\label{sec:guideline-DoInSimple-InternalTransition}

\begin{figure}[htbp]
  \centerline{\includegraphics[scale=\plantumlscale]{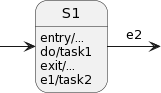}}
  \caption{A state with doActivity and internal transition.}
  \label{fig:DoInSimple-InternalTransition}
\end{figure}

\guidelineSubsect{Pattern}
In~\cref{fig:DoInSimple-InternalTransition} a simple state has an internal transition and a doActivity behavior (without an accept event action).

\guidelineSubsect{Possible intention}
A state machine can use internal transitions to react to incoming events without aborting the doActivity.
Contrary to self transitions, when firing an internal transition, the state is not exited and re-entered, thus exit and entry behavior is not executed, and the doActivity executing a long-running task is not interrupted.

\guidelineSubsect{Potential issues}
The execution of a doActivity might be concurrent with the firing of an internal transition.

\begin{warnEnum}
  \item\label{warn:concurrentInternalTransition}
  As there is no guarantee that actions are atomic,
  the execution of the \elementName{task1} behavior in the doActivity and \elementName{task2} effect in the internal transition can overlap.
\end{warnEnum}

\guidelineSubsect{Discussion}
\labelcref{warn:concurrentInternalTransition} might cause problems if
the tasks are not independent, e.g.,
they use shared variables,
send out signals that should not interleave, or
the internal transition effect expects that the execution of (some parts of) the doActivity has completed~(see \labelcref{warn:lateStart,warn:substateExpectsDoPartsFinished} later).

\subsection{DoActivity accepting events}
\label{sec:guidelineCategory-DoAccepting}

\guidelineSect{A doActivity with an accept event action}
\label{sec:guideline-DoAccepting-Basic}

\begin{figure}[htbp]
  \centering
  \subfloat[Basic case]{\includegraphics[scale=\plantumlscale]{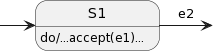}%
    \label{fig:DoAccepting-Basic}}
  \hfil
  \subfloat[Conflicting events]{\includegraphics[scale=\plantumlscale]{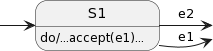}%
    \label{fig:DoAccepting-Basic-and-same-trigger}}
  \caption{A doActivity with an accept event action.}
  \label{fig:DoAccepting-Basic-subfigs}
\end{figure}

\guidelineSubsect{Pattern}
In~\cref{fig:DoAccepting-Basic} a simple state has a doActivity that includes an accept event action triggered by an event. The state has an outgoing transition triggered by another event. In~\cref{fig:DoAccepting-Basic-and-same-trigger} the same event triggers an outgoing transition.

\guidelineSubsect{Possible intention}
A doActivity can contain accept event actions, which can be used to react to external events, e.g., wait for a reply message during communication.

\guidelineSubsect{Potential issues}
In this case the doActivity's reactive behavior interferes with the reactivity of the state machine.

\begin{warnEnum}
  \item\label{warn:lateAccepterReg}
  In~\cref{fig:DoAccepting-Basic}, if the doActivity has not reached the accept event action by the time event \event{e1} is dispatched, the doActivity cannot accept the event, and the event is discarded due to the lack of a matching accepter.
  \item\label{warn:smCompeteForEvents}
  In~\cref{fig:DoAccepting-Basic-and-same-trigger} the situation is more complicated when both the state machine and the doActivity compete for the same event \event{e1}. Unlike transitions in the state machine where substates have priority, there is no priority defined between the state machine and its doActivities waiting for the same event. As a consequence, one of them will be non-deterministically chosen.
  \item\label{warn:abortWithoutWaitPoint}
  An explicit wait point might suggest that the doActivity can be aborted only when it is waiting for an event%
  \footnote{E.g., PSSM test case \pssmPageLink{83}{Behavior~003\=/A} seems to assume that abort only happens while waiting for events but test case \pssmPageLink{125}{Event~017\=/B} shows a possible abort before the doActivity execution could be observed.}.
  However, there is no guarantee that any action in the doActivity is completely executed before an outgoing transition (e.g., \event{e2}) aborts the doActivity. (See \labelcref{warn:lateStart,warn:abortEvenBeforeRun,warn:abortDuringAction}.)
\end{warnEnum}

\guidelineSubsect{Discussion}
In the non-conflicting case~(\cref{fig:DoAccepting-Basic}), using \inlineCmd{defer} to delay events for later processing in a doActivity can solve \cref{warn:lateAccepterReg}~(see \cref{sec:guideline-DoAccepting-Defer} for details).

In the conflicting case~(\cref{fig:DoAccepting-Basic-and-same-trigger}), the specification does not define priority between the state machine and the doActivity.
Hence, avoid using the same or overlapping triggering events in doActivities and the state machine.

\guidelineSect{A doActivity with an accept event action and the state defers the event}
\label{sec:guideline-DoAccepting-Defer}

\begin{figure}[htbp]
  \centering
  \subfloat[Basic case]{\includegraphics[scale=\plantumlscale]{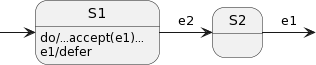}%
    \label{fig:DoAccepting-Defer}}
  \qquad
  \subfloat[Conflicting events]{\includegraphics[scale=\plantumlscale]{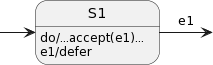}%
    \label{fig:DoAccepting-Defer-and-same-trigger}}
  \caption{An accept event action and deferring an event.}
  \label{fig:DoAccepting-Defer-and-same-trigger-subfigs}
\end{figure}

\guidelineSubsect{Pattern}
In~\cref{fig:DoAccepting-Defer} a simple state has a doActivity behavior that includes an accept event action triggered by an event \event{e1}. The state defers this event.
The state has an outgoing transition triggered by a different event, which targets state \stateSM{S2} with an outgoing transition on \event{e1}.
A special case is depicted in~\cref{fig:DoAccepting-Defer-and-same-trigger}, where state \stateSM{S1} has an outgoing transition on \event{e1} but also defers the same event.

\guidelineSubsect{Possible intention}
If there is no transition triggered by an event and the event would be discarded, \inlineCmd{defer} can be used to avoid discarding and delay the processing of the event to a later stage, e.g., to an upcoming state or to a later action in a doActivity.

\guidelineSubsect{Potential issues}
While defer ought to be a solution to harmonize the reactive behavior of the state machine and the doActivity, it comes with its own caveats.

\begin{warnEnum}
  \item\label{warn:deferredEventsHavePriorityAtAccept}
  While \stateSM{S1} is active it defers all incoming \event{e1} events.
  When a doActivity registers event accepters, it will first check the deferred event pool, even if an event with the same type is present in the normal event pool. These events can differ, e.g., if the events have parameters~(e.g., counters), and the doActivity is not guaranteed to get the latest event.%
  \item\label{warn:multipleEventsDeferred}
  While \stateSM{S1} is active, it defers \emph{all} incoming \event{e1} events. If the doActivity consumes only one or a few events, then exiting \stateSM{S1} releases the remaining deferred \event{e1} events. If next states (here~\stateSM{S2}) have transitions triggered by \event{e1}, the released \event{e1} events will be dispatched first in \stateSM{S2} instead of new incoming events.
  \item\label{warn:smCompeteForEventsDeferOverridingTransition}
  In the special case if events \event{e1} and \event{e2} are the same~(\cref{fig:DoAccepting-Defer-and-same-trigger}), the outgoing transition overrides the \inlineCmd{defer}, and we cannot ensure that the doActivity receives the event instead of the state machine. Before the doActivity registers an accepter, only the state machine can and will dispatch the event (similar to \labelcref{warn:lateAccepterReg}); after the registration they both compete for it (similar to \labelcref{warn:smCompeteForEvents}).
\end{warnEnum}

\guidelineSubsect{Discussion}
In the non-conflicting case~(\cref{fig:DoAccepting-Defer}), \inlineCmd{defer} is a solution to keep \event{e1} events dispatched in state \stateSM{S1} when the doActivity is not able to process it yet.
These events are delayed and not discarded.

However, in the conflicting case~(\cref{fig:DoAccepting-Defer-and-same-trigger}), \inlineCmd{defer} \emph{cannot be used} to prioritize between the state machine and its doActivities due to the overriding transition. The situation is even more complicated because priority depends on the exact timing of the event: before doActivity accepter registration the state machine accepts it; after the registration they compete for it.
Therefore, pay attention deferring an event when a state or its substate have an outgoing transition triggered by the same event, because that renders defer useless.

\inlineCmd{Defer} does not change how the doActivity runs, it only delays incoming events for a later accept. \labelcref{warn:lateStart,warn:abortEvenBeforeRun,warn:abortDuringAction} about the uncertainty when the doActivity starts and which actions are executed before the abort are still applicable here.

\subsection{DoActivity in a composite state}
\label{sec:guidelineCategory-DoInComposite}

\guidelineSect{DoActivity in a parent state}
\label{sec:guideline-DoInComposite-DoInParent}

\begin{figure}[htbp]
  \centerline{\includegraphics[scale=\plantumlscale]{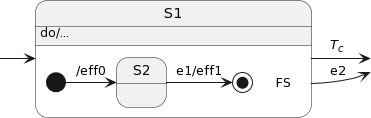}}
  \caption{A doActivity in a composite parent state.}
  \label{fig:DoInComposite-DoInParent}
\end{figure}

\guidelineSubsect{Pattern}
In~\cref{fig:DoInComposite-DoInParent} a doActivity is present in a composite state. The states have outgoing transitions, and the composite state has a completion transition.

\guidelineSubsect{Possible intention}
A composite state is a concise representation of the shared behavior of multiple states.
A long-running activity is executed in the doActivity but the system can continue its other tasks by handling external events in a more complex way than internal transitions~(\cref{sec:guideline-DoInSimple-InternalTransition}.

\guidelineSubsect{Potential issues}
While concurrency is usually modeled with orthogonal regions, doActivities alone can introduce concurrency for composite states which should be noted.
\begin{warnEnum}
  \item\label{warn:substateExpectsDoPartsFinished}
  The doActivity has its own separate lifecycle, and there are no guarantees about when it will start its execution or what parts are finished by the time a substate/transition behavior is executed. See \labelcref{warn:lateStart}. %
  \item\label{warn:concurrentSubstate}
  The doActivity can run concurrently with the state machine. The doActivity and the effect of transitions to/from substate \stateSM{S2} can concurrently execute steps observable from outside (e.g., sending signals) or can affect each other (e.g., via shared variables). See \labelcref{warn:concurrentInternalTransition}.
  \item\label{warn:completionNeedsRegionInFinalState}
  A composite state completes if \emph{all} of its internal activities have completed, i.e., entry and doActivity Behaviors, and all of its regions have reached a FinalState, so either the doActivity or the region can delay the firing of the completion transition \transition{T\textsubscript{c}}. %
\end{warnEnum}

\guidelineSubsect{Discussion}
Orthogonal regions clearly show concurrent behaviors but doActivities themselves without orthogonal regions can introduce concurrency for composite states which should be considered.
Care should be taken not to expect anything in the doActivity to be already performed by the time a nested transition is triggered. For example, initialization required by the substates should be put to separate states to ensure they are finished. This bad practice can be observed in the example in~\cref{fig:MeasuringComponent}. Also expect that the doActivity can overlap with any behavior in the region.

\guidelineSect{DoActivity in a substate}
\label{sec:guideline-DoInComposite-DoInSubstate}

\begin{figure}[htbp]
  \centerline{\includegraphics[scale=\plantumlscale]{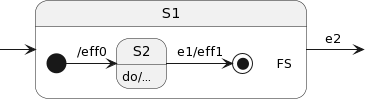}}
  \caption{A doActivity in a substate.}
  \label{fig:DoInComposite-DoInSubstate}
\end{figure}

\guidelineSubsect{Pattern}
In~\cref{fig:DoInComposite-DoInSubstate} a doActivity is present in a substate of a composite state. The states have outgoing transitions, and the composite state has a completion transition.

\guidelineSubsect{Potential issues}
This is very similar to \cref{sec:guideline-DoInSimple-NoAccept} (\labelcref{warn:lateStart,warn:abortEvenBeforeRun,warn:abortDuringAction}), the main difference is that not only direct outgoing transitions can abort the doActivity but transitions from the parent state too.
See also \labelcref{warn:completionNeedsRegionInFinalState}.

If the doActivity accepts event \event{e2}, then it will compete with the parent state's outgoing transition~(\labelcref{warn:smCompeteForEvents}). However, if \stateSM{S2} defers 
\event{e2}, then the doActivity should receive the event.

\guidelineSect{Multiple doActivities with accept event actions}
\label{sec:guideline-DoInComposite-MultipleDoAndAccept}

\guidelineSubsect{Pattern}
DoActivities are present in a composite state and also in its substate. The states have outgoing transitions.

\begin{figure}[htbp]
  \centerline{\includegraphics[scale=\plantumlscale]{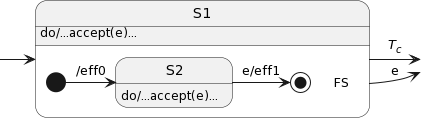}}
  \caption{Multiple doActivities with accept event actions.}
  \label{fig:DoInComposite-MultipleDoAndAccept}
\end{figure}

\guidelineSubsect{Potential issues}
As the state machine's behaviors and doActivities can run concurrently, more doActivities in a state machine can cause many more possible execution traces, all of which should be considered~(\labelcref{warn:concurrentSubstate}).
Both doActivities and the state machine can wait for the same event, in which case all doActivities and the state machine compete for it.
Unlike transitions from substates, which have priority, there is no priority defined between \stateSM{S1}'s doActivity and a transition from \stateSM{S2} substate, or the doActivity of \stateSM{S2} substate if they wait for the same event%
~(\labelcref{warn:smCompeteForEvents}).
Which one gets selected depends on the exact order of accepter registrations and event dispatching~(\labelcref{warn:lateAccepterReg}).
See also \labelcref{warn:substateExpectsDoPartsFinished,warn:completionNeedsRegionInFinalState}.

\guidelineSect{A doActivity with an accept event action and the composite state defers the event}
\label{sec:guideline-DoInComposite-DoAcceptAndDefer}

\begin{figure}[htbp]
  \centerline{\includegraphics[scale=\plantumlscale]{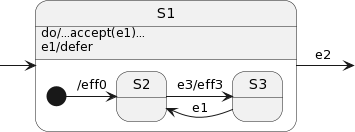}}
  \caption{DoActivity and composite state deferring events}
  \label{fig:DoInComposite-DoAcceptAndDefer}
\end{figure}

\guidelineSubsect{Pattern}
In~\cref{fig:DoInComposite-DoAcceptAndDefer} a composite state defers an event and its doActivity has an accept event action for it.

\guidelineSubsect{Possible intention}
A long-running activity is running in the doActivity while the substates handle other tasks and external events.
The doActivity waits for event \event{e1} at some point in its execution.
\inlineCmd{Defer} is used to delay the event until the doActivity can process it.

\guidelineSubsect{Potential issues} Deferring in a composite state presents new issues on top of existing ones (\labelcref{warn:smCompeteForEventsDeferOverridingTransition,warn:substateExpectsDoPartsFinished,warn:concurrentSubstate,warn:lateAccepterReg,warn:deferredEventsHavePriorityAtAccept}).

\begin{warnEnum}
  \item\label{warn:deferIsConfigSensitive} Whether the \stateSM{S1} state can defer the \event{e1} event (and thus the doActivity can consume it) changes depending on the current state configuration (i.e., it cannot defer in \stateSM{S1}[\stateSM{S3}] due to the higher-priority transition from \stateSM{S3}).
  \item\label{warn:stealDeferredEvents}
  If an event has been deferred, the doActivity can dispatch the event directly from the deferred event pool while the state machine is busy in an RTC step processing another event in a substate (e.g., \event{e3} in state \stateSM{S2}).
\end{warnEnum}

\guidelineSubsect{Discussion}
As whether the state machine can defer an event is re-evaluated upon every new dispatch, the result of defer can change between events depending on which substate is active. In \stateSM{S1}[\stateSM{S2}], the doActivity will get a new \event{e1}, but in \stateSM{S1}[\stateSM{S3}] it will compete with the state machine. 

Consider the other case (\labelcref{warn:stealDeferredEvents}) that the state machine has deferred an event \event{e1} because the doActivity has not reached its AcceptEventAction yet. If later the doActivity registers an accepter, then the doActivity ``steals'' the event from the deferred event pool even if the state machine is still in the middle of an RTC step and has not reached a stable state configuration, e.g., it is processing event \event{e3} in state \stateSM{S1}[\stateSM{S2}].

This behavior contradicts the UML clause that
an event is not dispatched while the state machine is busy processing the previous one~\umlPageCite{358}{14.2.3.9.1},
and the obvious expectation that
the event pool(s) are stable during RTC steps and only new events can be added to them.
This is rather surprising because when an RTC step is running, it is not clear whether the event is still to be deferred as in the source state or should be released back based on the new target state. 

Moreover, the state machine does not have such a ``shortcut'' (\labelcref{warn:stealDeferredEvents}); if it moves to \stateSM{S1}[\stateSM{S3}], it cannot accept a deferred \event{e1} event, only a newly dispatched one.

\subsection{Composite states and orthogonal regions with doActivities}
\label{sec:guidelineCategory-DoInCompositeAndOrthogonal}
\label{sec:guideline-DoInCompositeAndOrthogonal-}

\guidelineSubsect{Pattern}
There is a composite state with multiple orthogonal regions and doActivities at different hierarchy levels. DoActivities can accept events, and the composite state has a completion transition.

\begin{figure}[htbp]
  \centerline{\includegraphics[scale=\plantumlscale]{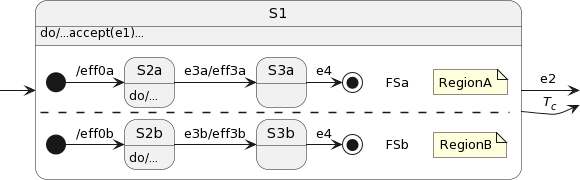}}
  \caption{Orthogonal regions and doActivities}
  \label{fig:DoInCompositeAndOrthogonal}
\end{figure}

\guidelineSubsect{Possible intention}
Composite states comprise shared behavior of the included states.
Orthogonal regions can be used to model different aspects of state machine behavior running concurrently.

\guidelineSubsect{Potential issues}
All concurrency and complexity that composite states, orthogonal regions and doActivities separately introduce can be present if we combine these modeling elements. Therefore we refer to issues mentioned above that are also applicable here.
Orthogonal regions have the advantage of explicitly showing the modeler's intention that the regions are concurrent with each other.
The modeler should consider all possible concurrency conflicts (e.g., between doActivities, entry/exit actions, (internal) transition effects) and the execution traces~(\labelcref{warn:concurrentSubstate}).
Event acceptance in the doActivity depends on the exact order of accepter registration and event dispatching~(\labelcref{warn:lateAccepterReg}).
See also \labelcref{warn:completionNeedsRegionInFinalState,warn:substateExpectsDoPartsFinished}.

\guidelineSubsect{Discussion}
Special care should be taken to consider all possible dependency, timing and concurrency issues, as even a small model can cause these kinds of issues.
This complexity cries out for tool support~\cite{10.1145/3579821,horvath-se-2023} that would help the modelers to highlight the possible concurrent behavior, exact conditions for event accepters, shared variables used in concurrent settings, etc.
Unfortunately such advanced tooling is still only scarcely available.

\begin{table*}[!t]
	
	\newcommand*{\catNameCell}[2]{\multirow{#1}{*}{\parbox{12mm}{#2}}}
	\newcommand*{\patternNameCell}[1]{#1}
	\newcommand*{\warnCatNameCell}[3]{\multicolumn{#1}{@{}c@{}#2}{#3}}
	
	\newcommand*{\warnColSep}{\hspace{5pt}}
	\newcommand*{\hlineVspace}{0ex}%
	\caption{DoActivity patterns and relevant issues.
		\emph{Notation:} issue is
		important~(\warnImportant{}),
		applicable~(\warnApplicable{}),
		slightly relevant~(\warnSlightly{}), and
		not relevant~(\warnNoText{})
		for the pattern.
	}
	\label{tab:warnings}
	\centering
	
	\begin{tabular}{lll|c@{\warnColSep}|c@{\warnColSep}c@{\warnColSep}c@{\warnColSep}|c@{\warnColSep}c@{\warnColSep}c@{\warnColSep}c@{\warnColSep}c@{\warnColSep}c@{\warnColSep}c@{\warnColSep}c@{\warnColSep}c@{\warnColSep}|c@{\warnColSep}c@{\warnColSep}c@{\warnColSep}c@{\warnColSep}c}
		\hline
		
& 	& \multicolumn{1}{r|}{Category:}	& \warnCatNameCell{1}{|}{start} &	\warnCatNameCell{3}{|}{executing} &	\warnCatNameCell{9}{|}{event handling} &	\warnCatNameCell{5}{}{finalization} \\

Sect.	& DoActivity	& \multicolumn{1}{r|}{Issue:}	& \labelcref{warn:lateStart}	& \labelcref{warn:concurrentInternalTransition}	& \labelcref{warn:substateExpectsDoPartsFinished}	& \labelcref{warn:concurrentSubstate}	& \labelcref{warn:loseEventWhileInDo}	& \labelcref{warn:SelfSignal-ReorderDelayLoseDuplicate}	& \labelcref{warn:lateAccepterReg}	& \labelcref{warn:smCompeteForEvents}	& \labelcref{warn:deferredEventsHavePriorityAtAccept}	& \labelcref{warn:multipleEventsDeferred}	& \labelcref{warn:smCompeteForEventsDeferOverridingTransition}	& \labelcref{warn:deferIsConfigSensitive}	& \labelcref{warn:stealDeferredEvents}	& \labelcref{warn:abortEvenBeforeRun}	& \labelcref{warn:abortDuringAction}	& \labelcref{warn:doActivityForever}	& \labelcref{warn:abortWithoutWaitPoint}	& \labelcref{warn:completionNeedsRegionInFinalState}	\\%
[\hlineVspace]\hline \labelcref{sec:guideline-DoInSimple-NoAccept}	& \catNameCell{4}{in\\simple state}	& \patternNameCell{basic case}	& \warnImportant	& \warnNo	& \warnNo	& \warnNo	& \warnNo	& \warnNo	& \warnNo	& \warnNo	& \warnNo	& \warnNo	& \warnNo	& \warnNo	& \warnNo	& \warnImportant	& \warnImportant	& \warnNo	& \warnNo	& \warnNo	\\%
\labelcref{sec:guideline-DoInSimple-NoAccept-Completion}	& 	& \patternNameCell{with completion transition}	& \warnNo	& \warnNo	& \warnNo	& \warnNo	& \warnImportant	& \warnNo	& \warnNo	& \warnNo	& \warnSlightly	& \warnSlightly	& \warnNo	& \warnNo	& \warnNo	& \warnSlightly	& \warnSlightly	& \warnImportant	& \warnNo	& \warnNo	\\%
\labelcref{sec:guideline-DoInSimple-SelfSignaling}	& 	& \patternNameCell{with self-signaling}	& \warnNo	& \warnNo	& \warnNo	& \warnNo	& \warnNo	& \warnImportant	& \warnNo	& \warnNo	& \warnSlightly	& \warnSlightly	& \warnNo	& \warnNo	& \warnNo	& \warnSlightly	& \warnSlightly	& \warnNo	& \warnNo	& \warnNo	\\%
\labelcref{sec:guideline-DoInSimple-InternalTransition}	& 	& \patternNameCell{with internal transition}	& \warnApplicable	& \warnImportant	& \warnApplicable	& \warnNo	& \warnNo	& \warnNo	& \warnNo	& \warnNo	& \warnNo	& \warnNo	& \warnNo	& \warnNo	& \warnNo	& \warnSlightly	& \warnSlightly	& \warnNo	& \warnNo	& \warnNo	\\%
[\hlineVspace]\hline \labelcref{sec:guideline-DoAccepting-Basic}	& \catNameCell{2}{accepting events}	& \patternNameCell{basic case}	& \warnApplicable	& \warnNo	& \warnNo	& \warnNo	& \warnNo	& \warnNo	& \warnImportant	& \warnImportant	& \warnNo	& \warnNo	& \warnNo	& \warnNo	& \warnNo	& \warnApplicable	& \warnApplicable	& \warnNo	& \warnImportant	& \warnNo	\\%
\labelcref{sec:guideline-DoAccepting-Defer}	& 	& \patternNameCell{state defers the event}	& \warnApplicable	& \warnNo	& \warnNo	& \warnNo	& \warnNo	& \warnNo	& \warnApplicable	& \warnApplicable	& \warnImportant	& \warnImportant	& \warnImportant	& \warnSlightly	& \warnNo	& \warnApplicable	& \warnApplicable	& \warnNo	& \warnSlightly	& \warnNo	\\%
[\hlineVspace]\hline \labelcref{sec:guideline-DoInComposite-DoInParent}	& \catNameCell{4}{in\\composite state}	& \patternNameCell{doActivity in a parent state}	& \warnApplicable	& \warnApplicable	& \warnImportant	& \warnImportant	& \warnNo	& \warnNo	& \warnSlightly	& \warnSlightly	& \warnNo	& \warnNo	& \warnNo	& \warnNo	& \warnNo	& \warnSlightly	& \warnSlightly	& \warnNo	& \warnSlightly	& \warnImportant	\\%
\labelcref{sec:guideline-DoInComposite-DoInSubstate}	& 	& \patternNameCell{doActivity in a substate}	& \warnApplicable	& \warnNo	& \warnNo	& \warnNo	& \warnNo	& \warnNo	& \warnSlightly	& \warnApplicable	& \warnNo	& \warnNo	& \warnNo	& \warnNo	& \warnNo	& \warnApplicable	& \warnApplicable	& \warnNo	& \warnSlightly	& \warnSlightly	\\%
\labelcref{sec:guideline-DoInComposite-MultipleDoAndAccept}	& 	& \patternNameCell{multiple doActivities accepting events}	& \warnSlightly	& \warnNo	& \warnApplicable	& \warnApplicable	& \warnNo	& \warnNo	& \warnApplicable	& \warnApplicable	& \warnNo	& \warnNo	& \warnNo	& \warnNo	& \warnNo	& \warnSlightly	& \warnSlightly	& \warnNo	& \warnSlightly	& \warnApplicable	\\%
\labelcref{sec:guideline-DoInComposite-DoAcceptAndDefer}	& 	& \patternNameCell{accepting events and state defers event}	& \warnSlightly	& \warnNo	& \warnApplicable	& \warnApplicable	& \warnNo	& \warnNo	& \warnApplicable	& \warnSlightly	& \warnApplicable	& \warnSlightly	& \warnApplicable	& \warnImportant	& \warnImportant	& \warnSlightly	& \warnSlightly	& \warnNo	& \warnSlightly	& \warnSlightly	\\%
[\hlineVspace]\hline \labelcref{sec:guideline-DoInCompositeAndOrthogonal-}	& \multicolumn{2}{l|}{in composite states and orthogonal regions}	& \warnSlightly	& \warnNo	& \warnApplicable	& \warnApplicable	& \warnNo	& \warnNo	& \warnApplicable	& \warnSlightly	& \warnNo	& \warnNo	& \warnNo	& \warnNo	& \warnNo	& \warnSlightly	& \warnSlightly	& \warnNo	& \warnSlightly	& \warnApplicable
		
		\\%
		[\hlineVspace]\hline
	\end{tabular}
\end{table*}

\subsection{Pattern and issue categories}

This section summarizes the doActivity patterns and issues mentioned above.
\Cref{tab:warnings} shows which issue is relevant to which pattern. The issues are grouped into categories based on the phases of the doActivity, as examined in~\cref{sec:deep-dive}:

\begin{itemize}
  \item \textbf{Start}~[\cref{sec:deepDive:starting}]:
        \labelcref{warn:lateStart} points out that when the doActivity actually starts its execution is uncertain due to its separate lifecycle.
  \item \textbf{Executing}~[\cref{sec:deepDive:executing}]:
        \labelcref{warn:concurrentInternalTransition,warn:substateExpectsDoPartsFinished,warn:concurrentSubstate} call attention to activities concurrent with doActivities (e.g., internal transitions, substates and their transitions, orthogonal regions), which might cause problems, e.g., if they use the same variables, send out interleaving signals, or depend on potentially unfinished parts of the doActivity.
  \item \textbf{Event handling}~[\cref{sec:deepDive:events}]:
        \labelcref{warn:lateAccepterReg,warn:smCompeteForEvents,warn:loseEventWhileInDo} show that the timing and priority of event accepters in the state machine and the doActivities are complex.
        \labelcref{warn:SelfSignal-ReorderDelayLoseDuplicate} describes problems with signals in the case of self-signaling.
        \labelcref{warn:multipleEventsDeferred,warn:deferredEventsHavePriorityAtAccept,warn:smCompeteForEventsDeferOverridingTransition,warn:stealDeferredEvents,warn:deferIsConfigSensitive} draw attention to event deferral, which should be used with care:
        it depends on state configuration and accepters;
        deferred events have priority over normal ones both in a doActivity and when they are no longer deferred;
        a doActivity can dispatch a deferred event while the state machine is busy executing an RTC step; and
        transitions can override \inlineCmd{defer}.
  \item \textbf{Finalization}~[\cref{sec:deepDive:finalization}]:
        \labelcref{warn:doActivityForever,warn:abortEvenBeforeRun,warn:abortDuringAction,warn:abortWithoutWaitPoint} show that the doActivity can run without limits until exiting the state aborts it. Abort might happen before executing any action or even during an action.
        \labelcref{warn:completionNeedsRegionInFinalState} emphasizes the conditions needed for a state to complete.
\end{itemize}

We investigated how these patterns are used in practice in the context of the TMT model. We found all the proposed patterns in TMT with the exception of placing a doActivity in a composite parent state~(\labelcref{sec:guideline-DoInComposite-DoInParent,sec:guideline-DoInComposite-MultipleDoAndAccept,sec:guideline-DoInComposite-DoAcceptAndDefer}).
The supplementary material includes a detailed mapping table.

\section{Related work}\label{sec:related-work}

\subsection{Usage of doActivities in UML state machines}

Although doActivity behaviors are mentioned in most textbooks, limited public information is available about their usage in engineering practice. We analyzed public datasets about UML usage, but found them less relevant. According to our experience originating from industrial R\&D projects, doActivities become more important in complex, executable models typical in embedded or safety-critical domains. 

Most publicly available UML artifacts are for simpler models mainly used for high-level documentation, in which doActivities and other complex language constructs are not used. For example, Langer et al.~\cite{mci/Langer2014} present a study on the usage of UML sublanguages and model elements in 150 Enterprise Architect models collected using Google Search; but state machines are not detailed. To find further real-world UML state machine models, we analyzed the Lindholmen Dataset~\cite{DBLP:conf/models/HebigHCRF16}, the largest still available dataset containing about 93 thousand UML files from 24 thousand GitHub repositories.
We used the ``state'' search keyword for file paths and found 2635 file versions from 734 GitHub repositories. %
We filtered on UML images and models containing complex constructs (e.g., entry/exit actions, doActivities, composite states),
and we manually inspected the remaining 96 images and 288 XMI/UML files more thoroughly.
We excluded repositories of UML modeling tools since they contain mostly UML metamodels, test and sample models.
The remaining models mainly were home assignments for university courses, sample UML models, and there were some illustrative diagrams to document projects.
None of them was a system designed using UML state machines.

The authors of the Lindholmen dataset also warn that GitHub contains a high amount of student and toy repositories. This warning confirms our experience that open repositories are not the preferred choice for storing complex design models, and mining such repositories cannot be used to estimate the usage of complex UML elements realistically.

A more realistic source of information about industry practices is, for example, the guidelines and patterns of the OpenSE Cookbook~\cite{karban2018opense}, which collects the experiences of numerous model-based systems engineering projects from NASA JPL and ESO. A significantly extended, v2 version of the cookbook is available online~\cite{OpenSE-CookbookV2}. The cookbook shows patterns of how doActivities are used to model long-running tasks in simulation and duration analysis, or how self-signaling and completion transitions combine activities and state machines. Moreover, the cookbook presents complex, real-world examples from the TMT model~\cite{url-tmt} (e.g., it is worth looking at the Procedure Executive and Analysis Software state machine containing numerous doActivities).

\subsection{Guidelines and patterns for UML state machines}

Several style guides and validation rules are available for UML state machines, but they do not provide detailed patterns for doActivities. Ambler~\cite{uml2sytle-book} provides useful but high-level advice for arranging and naming UML elements. Torre et al.~\cite{TORRE2018121} collected 687 consistency rules for various UML diagrams. The SAIC Digital Engineering Validation Tool\footnote{\url{https://www.saic.com/digital-engineering-validation-tool}.} is one of the most extensive public validation rule sets. However, none of the rules mention doActivities. 

Das and Dingel~\cite{DasD18} collected conventions, patterns and antipatterns for UML-RT. Similarly, they provided consequences and possible refactorings for each pattern. But as UML-RT does not have doActivities, our results are orthogonal. Alenazi et al.~\cite{Alenazi2019} collected 42 mistakes mentioned in studies of SysML, but they also do not cover doActivities.

\subsection{Semantics for UML state machines}

\textbf{Execution and simulation}
Ciccozzi et al.~\cite{Ciccozzi2018} provide a survey about execution approaches and tools for UML models. Most of the approaches support class, state machine and activity diagrams. However, 13 of the 82 solutions are based on the precise fUML semantics, and the survey does not mention PSSM. 
Micskei et al.~\cite{micskei2014oss4mde-2014} collected available tools and experiences about executable UML, specifically about fUML activities and Alf action language. 
Guermazi et al.~\cite{guermazi:cea-01844057} report the lessons learned while implementing fUML and Alf modeling in Eclipse Papyrus and the Moka execution plug-in.
Regarding the representation of possible execution traces~(\cref{fig:smMicroStepsOrthogonalRegionConstraints}), most approaches do not detail the exact steps while firing a complex transition.
The closest approach was by Pintér and Majzik~\cite{Pinter2007}, who used PERT graphs to express the execution dependencies between basic steps. 

\textbf{Code generation}
DoActivity semantics is also interesting from a code generation perspective.
A systematic review of papers between 1992--2010 performed by Dom{\'{\i}}nguez et al.~\cite{code-generation-slr} shows that only a minority of the approaches support concurrency and even fewer support doActivities to a minimum extent. 
Metz et al.~\cite{metz1999code} introduce code generation concepts, but only for UML 1.1 statechart diagrams, including interruptable doActivities that run on separate threads and are interrupted by incoming events, automatic (completion) transitions, and also mention deferred events. They define modeling constraints, e.g., proposing interruptable activities to be modeled as doActivities and non-interruptable activities as entry behaviors, which are atomic in their work.

Pham et al.~\cite{DBLP:conf/modelsward/PhamRGL17a} provide C++ code generation for UML composite structures and state machines in Papyrus.
The approach supports concurrency, including doActivities on separate threads, RTC steps, and an event pool with completion and deferred events. Event accepters are not mentioned.
They compared the Moka simulation of 66 PSSM test cases and results from their tool to verify the semantics. However, it is not detailed how they produced the same execution trace both in Moka and their tool for concurrent behaviors where multiple possible traces exist.

Umple~\cite{DBLP:journals/scp/LethbridgeFBBGA21, alghamdi2015umple} is a software modeling tool and textual programming language. It supports combining UML class and state diagrams with many object-oriented languages, and generating code from them. The tool supports concurrent regions, entry/exit actions, concurrent doActivities.
As Umple aims for the co-existence of code and models, it alters the semantics of UML state machines to enable efficient code generation (e.g., there can be no conflicting transitions on the same level, transitions in orthogonal regions fire after each other sequentially in a fixed order, there is no native way to accept events in doActivities).
In Umple a single component can have multiple state machines, which run after each other and process events from the same event pool.
Umple offers \emph{pooled} state machines as a simplified version of \inlineCmd{defer} to keep events in the pool if there is no outgoing transition to process it.
Thus, many of the non-determinism of the UML/PSSM are not present in Umple, and only
\cref{warn:lateStart,warn:abortEvenBeforeRun,warn:doActivityForever,warn:multipleEventsDeferred,warn:substateExpectsDoPartsFinished,warn:concurrentSubstate,%
warn:abortDuringAction,warn:loseEventWhileInDo}
are relevant for Umple.

\textbf{Verification}
Andr\'{e} et al.~\cite{10.1145/3579821} provide a survey of UML state machine semantics formalizations for model checking purposes from 1997 to 2021. They collected 45 translation-based and 16 operational semantics, and categorized whether they support the 17 non-trivial UML features chosen. 
We examined the relevant UML 2.x papers where either doActivity support was explicitly mentioned or advanced elements are supported (RTC, orthogonal regions).

Fecher and Sch{\"{o}}nborn~\cite{DBLP:conf/fmics/FecherS06} give a translation from UML state machines into the ``core state machines'' semantic domain using natural language. Although it is one of the most complete approaches, it does not define RTC steps properly and does not support entry/exit behaviors.
André et al.~\cite{DBLP:journals/fac/AndreBC16} propose a formalization using colored Petri nets~(CPN) that supports RTC steps and certain concurrency aspects without support for deferred events, and only simple states can have doActivities. The semantics is non-standard because a doActivity can be executed as often as wished. 
Liu et al.~\cite{DBLP:conf/ifm/LiuLACSWD13} propose a formal operational semantics using Labeled Transition System~(LTS) for all features of UML version 2.4.1 (except time events) and implemented their approach in the USMMC model checker.
The semantics support deferred events and doActivities. The doActivities are in real concurrency with other behaviors of the containing state, however, how event acceptance works in doActivities is not mentioned.
Due to unclarities in UML they restrict the possible executions of (compound) transitions in orthogonal regions by treating certain parts as atomic %
instead of executing them completely in parallel as PSSM does. This restriction also applies to entry/exit behaviors.
The limitation of their work comes from the assumptions made to resolve UML unclarities, and the lack of formally defined action language, which would be useful for the definition and analysis of entry/do/exit behaviors and transition effects.

Abdelhalim et al.~\cite{DBLP:journals/sttt/AbdelhalimST13} formalize UML activity diagrams using Communicating Sequential Processes~(CSP) for model checking. They model the event pool and waiting event accepters following the fUML standard. However they do not consider activities in the context of state machines.

\subsection{DoActivities in other state machine variants}

Some other state machine variants also have doActivities or similar features.
Harel's statecharts~\cite{DBLP:journals/scp/Harel87} from 1987 already had a doActivity-like feature: an activity can be carried out continuously \emph{throughout} the system being in a specific state, i.e., it is started on entering and stopped on leaving that state. In the original version, most of the issues discussed were not present since there were no detailed executable activity definitions.
Other variants like UML-RT, a UML profile for real-time systems, do not support doActivities. 

In SCXML~\cite{barnett2015scxml}, the closest construct to doActivity is \emph{invoke}, which is used to create an instance of an external service. The invoked service can be interrupted %
if the containing state is exited,%
and the state is left only when the service is completely stopped.
The state machine and the service communicate by sending events. We are not aware of works focusing on invoke-related concurrency issues.

Crane and Dingel~\cite{DBLP:journals/sosym/CraneD07} compare Harel, UML and Rhapsody state machines, and emphasize that modelers should be aware of different interpretations of models in various formalisms, especially for different execution behaviors.

\section{Conclusion}\label{sec:conclusion}

This paper presented the use of doActivity behaviors in UML state machines and made two contributions. 1) We examined the operational semantics defined in the PSSM specification, and based on an analysis of available artifacts, we synthesized several insights about the complex interplay of doActivity and state machine semantics. Some of these insights are not mentioned anywhere in the description or conformance tests of the specification and, to the best of our knowledge, have not been reported in the literature and are unknown to most engineers (e.g., priority of accepters). These observations might seem to be minor details, but they can introduce serious non-deterministic errors that are especially hard to detect or debug. 2) Based on the semantic insights, we systematically collected 11 patterns of using doActivities, highlighted things to consider in each situation, and recommended countermeasures to potential issues. The descriptions of the patterns are practical, and we hope they offer actionable advice to engineers without requiring deep knowledge of the operational semantics.

Such semantic details are essential when state machines are used for detailed, executable design. 
DoActivities are inherently concurrent and non-deterministic elements. Numerous alternative traces shall be considered and reasoned about when doActivities are used even in simple state machines. If the engineers who develop or implement these models are unaware of all the possible behaviors, or tools working with the models do not conform to the specification, then the conflicting understanding of the semantics could lead to numerous problems~\cite{conf/ecmfa2018/baduel}. Our patterns and guidelines aim to mitigate these issues.

As future directions, we plan to conduct a case study on employing the patterns in an industrial setting and will analyze the redesigned semantics of the new SysMLv2 language \cite{journal/sanford/sysmlv2} to make recommendations regarding do actions, contributing to the Execution Annex of KerML~\cite{omg-kerml}.

\ifCLASSOPTIONcompsoc
\section*{Acknowledgments}
\else
\section*{Acknowledgment}
\fi

This work was supported by project no. 2019-1.3.1-KK-2019-00004 under the 2019-1.3.1-KK funding scheme, project no. 2022-1.2.4-EUREKA-2023-00013 under the 2022-1.2.4-EUREKA funding scheme, and the ÚNKP-23-3-II-BME-362 New National Excellence Program of the Ministry for Culture and Innovation from the source of the National Research, Development and Innovation Fund of Hungary.

\ifCLASSOPTIONcaptionsoff
  \newpage
\fi

\bibliographystyle{IEEEtran}

\vskip -2\baselineskip plus -1fil
\begin{IEEEbiography}[{\includegraphics[width=1in,height=1in,clip,keepaspectratio]{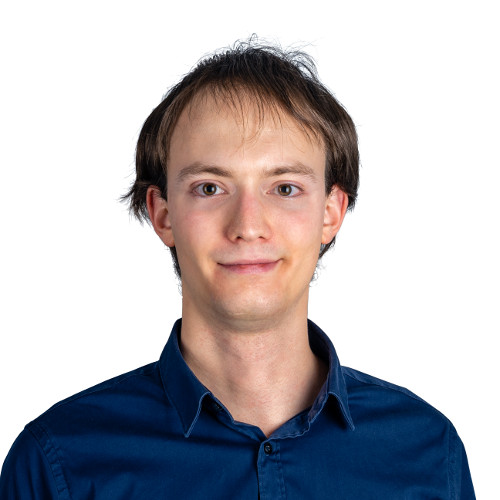}}]{Márton Elekes}
  is a PhD candidate at the Budapest University of Technology and Economics. His research interests include software testing and graph databases. He participates in R\&D projects about testing service-oriented systems and dataspaces. He was a contributor of the LDBC Social Network Benchmark. He participates in the Conformance Working Group of OMG's Systems Modeling Community.
\end{IEEEbiography}
\vskip -3\baselineskip plus -1fil
\begin{IEEEbiography}[{\includegraphics[width=1in,height=1in,clip,keepaspectratio]{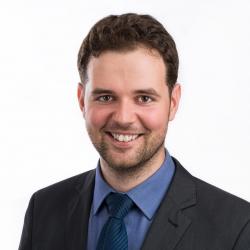}}]{Vince Molnár}
  is an assistant professor at the Budapest University of Technology and Economics. His main research field is model-based development and formal methods. He is leading the development of the Gamma Framework. He was a member of the SysML v2 Submission Team and is leading the Conformance and Formal Methods Working Groups of OMG's Systems Modeling Community.
\end{IEEEbiography}
\vskip -3\baselineskip plus -1fil
\begin{IEEEbiography}[{\includegraphics[width=1in,height=1in,clip,keepaspectratio]{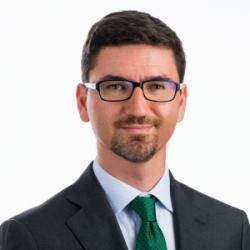}}]{Zoltán Micskei}
  received the Ph.D. degree from the Budapest University of Technology and Economics. He is currently an associate professor at the same university, leading the Critical Systems Research Group. His research interests include software testing and model-based engineering with a focus on empirical studies. He is a member of the Hungarian Young Academy, and a Senior Member of the ACM.
\end{IEEEbiography}

\end{document}